\documentclass[aps,prd,showpacs,notitlepage,nofootinbib,preprintnumbers,amsmath,amssymb]{revtex4-1}

\allowdisplaybreaks

\usepackage{graphics,graphicx}% Include figure files
\usepackage{dcolumn}% Align table columns on decimal point
\usepackage{bm}% bold math
\usepackage{epsfig}
\usepackage[usenames]{color}
\usepackage[colorlinks=true,citecolor=blue,linkcolor=blue,urlcolor=blue]{hyperref}
\usepackage{mathbbol}
\usepackage{epstopdf}
\usepackage{simplewick}
\usepackage[utf8]{inputenc} 
\usepackage{slashed}
\usepackage{stackrel}
\usepackage{mathrsfs}
\usepackage{xcolor}
\usepackage{subcaption}
\usepackage[most]{tcolorbox}
\tcbset{
  top=-5pt,
}

\usepackage{hyperref, empheq}
\usepackage[offset=1.5em]{simpler-wick}
\usepackage{xcolor, graphicx, slashed,  accents,  multirow,caption, subcaption}
\usepackage[most]{tcolorbox}

\usepackage{ragged2e} 
\DeclareCaptionJustification{justified}{\justifying}

\def\eq#1{{Eq.~(\ref{#1})}}
\def\eqs#1{{Eqs.~(\ref{#1})}}
\def\fig#1{{Fig.~\ref{#1}}}
\newcommand{\ben}{\begin{eqnarray*}}
\newcommand{\een}{\end{eqnarray*}}
\newcommand{\un}[1]{\underline{#1}}

\newcommand{\pd}{\partial}

\newcommand{\tr}{\mbox{tr}}
\newcommand{\thalf}{\tfrac{1}{2}}
\newcommand{\llangle}{\Big\langle \!\! \Big\langle}
\newcommand{\rrangle}{\Big\rangle \!\! \Big\rangle}
\newcommand{\cc}{\mbox{c.c.}}

\newcommand{\as}{\alpha_s}

\def\eq#1{{Eq.~(\ref{#1})}}
\def\eqs#1{{Eqs.~(\ref{#1})}}
\def\fig#1{{Fig.~\ref{#1}}}

\newcommand{\dhd}{{\textstyle d}
\lower.03ex\hbox{\kern-0.38em$^{\scriptstyle-}$}\kern-0.05em{}}
\newcommand{\dbar}{{\textstyle \delta}
\lower.03ex\hbox{\kern-0.38em$^{\scriptstyle-}$}\kern-0.05em{}}
\newcommand{\half}{{1\over 2}}

\newcommand{\ul}[1]{\underline{#1}}

\newcommand{\tord}{\textrm{T} \:}
\newcommand{\atord}{\bar{\textrm{T}} \:}

\makeatletter
\DeclareRobustCommand{\cev}[1]{%
  {\mathpalette\do@cev{#1}}%
}
\newcommand{\do@cev}[2]{%
  \vbox{\offinterlineskip
    \sbox\z@{$\m@th#1 x$}%
    \ialign{##\cr
      \hidewidth\reflectbox{$\m@th#1\vec{}\mkern4mu$}\hidewidth\cr
      \noalign{\kern-\ht\z@}
      $\m@th#1#2$\cr
    }%
  }%
}
\makeatother

\begin{document}

\title{Probing Orbital Angular Momentum Distributions in Elastic Vector Meson Production in Longitudinally Polarized Electron--Proton Scattering}

\author{Yuri V. Kovchegov}
  \email[Email: ]{kovchegov.1@osu.edu}
	\affiliation{Department of Physics, The Ohio State
           University, Columbus, OH 43210, USA}
           
\author{Brandon Manley}
\email[Email: ]{brandon.manley@uconn.edu}
\affiliation{Department of Physics, The Ohio State University, Columbus, Ohio 43210, USA}
\affiliation{Department of Physics, University of Connecticut, Storrs, Connecticut 06269, USA}

\begin{abstract}
\noindent 
We consider elastic vector meson production in longitudinally polarized electron-proton collisions at small $x$. Inspired by previous pioneering proposals in the literature, we calculate the numerator of the longitudinal double-spin asymmetry (DSA) within the polarized small-$x$ formalism for both the transversely and longitudinally polarized vector mesons. Our calculation demonstrates that the $\cos\phi_{k\Delta}$ azimuthal harmonic of the DSA for the elastic vector meson production at small transverse momentum transfer (with $\phi_{k\Delta}$ the azimuthal angle between the transverse momenta of the outgoing electron and meson in the dipole/Trento frame) provides a promising probe of the orbital angular momentum (OAM) distributions at small $x$. Our results can be applied to several different vector meson species, thus opening up a new class of observables that can be used to extract OAM distributions from experimental data. 
\end{abstract}
\maketitle

\tableofcontents

\section{Introduction}

Out of all the contributions to the proton spin budget, the orbital angular momenta (OAM) of the quarks and gluons are the only ones that have never been extracted from experiments. The OAM contribution to the proton spin can be represented using the Jaffe--Manohar sum rule \cite{Jaffe:1989jz} (see also the Ji sum rule \cite{Ji:1996ek}), which reads
\begin{align} \label{JM}
S_{q + \bar q} + L_{q + \bar q} + S_G + L_G = \frac{1}{2} 
\end{align}
with the quark and gluon spin contributions $S_{q + \bar q} (Q^2)$ and $S_G (Q^2)$ given by the integrals of the helicity parton distribution functions (hPDFs), and the OAM contributions given by integrals over the OAM distributions \cite{Bashinsky:1998if, Hagler:1998kg, Harindranath:1998ve, Hatta:2012cs, Ji:2012ba}
\begin{align} \label{oam_jm}
L_{q + \bar q} (Q^2) = \int\limits_0^1 dx \, L_{q + \bar q} (x, Q^2) , \qquad\qquad L_G (Q^2) = \int\limits_0^1 dx \, L_G (x, Q^2) .
\end{align}
Here $x$ is the parton momentum fraction and $Q$ is the momentum scale. Decades of polarized deep inelastic scattering (DIS), semi-inclusive DIS (SIDIS), and polarized proton-proton scattering measurements have led to constraints on the hPDFs (see \cite{EuropeanMuon:1987isl, Boer:2011fh, Aidala:2012mv, Accardi:2012qut, Leader:2013jra, Aschenauer:2013woa, Aschenauer:2015eha, Proceedings:2020eah, Ji:2020ena, AbdulKhalek:2021gbh} and references therein), while the OAM distributions $L_{q + \bar q} (x,Q^2)$ and $L_G (x,Q^2)$ remain phenomenologically unknown. Their $Q^2$ evolution has been known at one loop for some time \cite{Hagler:1998kg, Hoodbhoy:1998yb, Hatta:2019csj}, but it has never been put to phenomenological use, simply because no observable was known to couple to OAM.

Over the last decade, this situation has begun to change. A number of exclusive processes in electron--proton scattering have been identified 
as OAM probes in \cite{Ji:2016jgn, Hatta:2016aoc, Bhattacharya:2022vvo, Bhattacharya:2023hbq, Bhattacharya:2024sck, Kovchegov:2024wjs, Bhattacharya:2026qnd}, among them elastic dijet production, exclusive pion production, and elastic vector meson production. 

The mechanism they have in common relies on the fact that in order to measure the $z$-component of the OAM, $\vec L = \vec b \times \vec p$, one needs to measure the impact parameter $b_\perp$ between the parton and the proton. In turn, $b_\perp$ is a Fourier conjugate of  $\Delta_\perp$, the transverse momentum transfer, which can only be measured in exclusive processes. Measuring the resulting $\Delta_\perp$-dependent azimuthal correlation is what gives one access to the OAM distributions.

A particularly important region of the proton spin budget is the small-$x$ region of the integrals in \eq{oam_jm} (and the associated integrals for $S_{q+\bar{q}}$ and $S_G$). There, despite the fact that the hPDFs and OAM distributions are sub-eikonal (suppressed by one power of $x$) quantities, small-$x$ evolution produces rapidly growing partonic densities which may be an important source of proton spin. In recent years the $s$-channel/shock-wave formalism \cite{Mueller:1994rr, Mueller:1994jq, Mueller:1995gb, Balitsky:1995ub, Balitsky:1998ya, Kovchegov:1999yj, Kovchegov:1999ua, Jalilian-Marian:1997dw, Jalilian-Marian:1997gr, Weigert:2000gi, Iancu:2001ad, Iancu:2000hn, Ferreiro:2001qy} (see \cite{Gribov:1984tu, Iancu:2003xm, Weigert:2005us, JalilianMarian:2005jf, Gelis:2010nm, Albacete:2014fwa, Kovchegov:2012mbw, Morreale:2021pnn} for reviews) has been extended to sub-eikonal accuracy \cite{Altinoluk:2014oxa, Balitsky:2015qba, Balitsky:2016dgz, Kovchegov:2017lsr, Kovchegov:2018znm, Chirilli:2018kkw, Jalilian-Marian:2018iui, Jalilian-Marian:2019kaf, Altinoluk:2020oyd, Boussarie:2020vzf, Boussarie:2020fpb, Kovchegov:2021iyc, Altinoluk:2021lvu, Kovchegov:2022kyy, Altinoluk:2022jkk, Altinoluk:2023qfr, Altinoluk:2023dww, Li:2023tlw}, with the small-$x$ contribution to the proton spin serving as one of the main applications of these developments. We refer to the version of this formalism used below as the light-cone operator treatment (LCOT). Within it, the hPDFs at small $x$ are expressed through polarized dipole amplitudes obeying closed evolution equations in the large-$N_c$ \cite{tHooft:1973alw} and large-$N_c \& N_f$ \cite{Veneziano:1976wm} limits \cite{Kovchegov:2015pbl, Kovchegov:2016zex, Kovchegov:2016weo, Kovchegov:2017jxc, Kovchegov:2017lsr, Kovchegov:2018znm, Cougoulic:2019aja, Kovchegov:2020hgb, Cougoulic:2020tbc, Chirilli:2021lif, Kovchegov:2021lvz, Cougoulic:2022gbk, Borden:2023ugd, Adamiak:2023okq, Borden:2024bxa}. (Here, $N_c$ and $N_f$ are the numbers of quark colors and flavors.) The LCOT formalism has been used to successfully describe world scattering data from polarized deep inelastic scattering (DIS), semi-inclusive DIS (SIDIS), and single-inclusive jet production in proton-proton collisions \cite{Adamiak:2021ppq, Adamiak:2023yhz, JAMCollaborationSmall-xAnalysisGroup:2025tfa}. The OAM distributions require one further ingredient. As shown in \cite{Kovchegov:2019rrz, Kovchegov:2023yzd, Becker:2026xcp}, to describe the OAM distributions, one further requires the first impact-parameter moments of the polarized dipole amplitudes, which we refer to as the moment amplitudes. Small-$x$ evolution for the moment amplitudes was derived and numerically solved at large $N_c$ in \cite{Kovchegov:2023yzd}, analytically solved at large $N_c$ in \cite{Manley:2024pcl}, and recently extended to large $N_c \& N_f$ in \cite{Becker:2026xcp}, where the sub-eikonal $F^{+-}$ operator \cite{Chirilli:2018kkw, Chirilli:2021lif, Altinoluk:2021lvu, Li:2023tlw, Kovchegov:2025gcg} was shown to mix with the previously known operators. Any observable proposed as an OAM probe at small $x$ must therefore be sensitive to the moment amplitudes. For an elastic process, the transverse momentum transfer $\Delta_\perp$ is conjugate to the impact parameter of the dipole--proton scattering. Elastic observables at linear order in $\Delta_\perp$ provide access to the moment amplitudes and therefore the OAM distributions.

Elastic dijet production in longitudinally polarized electron--proton collisions at small $x$ was analyzed along these lines in our earlier work \cite{Kovchegov:2024wjs} (see \cite{Altinoluk:2022jkk, Agostini:2024xqs, Altinoluk:2024zom, Altinoluk:2024tyx, Agostini:2025vvx} for related calculations), with the $F^{+-}$ contribution to the process added in \cite{Becker:2026xcp}. In the present work, we consider another channel which has been first proposed in \cite{Bhattacharya:2026qnd}, namely elastic vector meson production, $\gamma^* + p \to V + p$. Elastic vector meson production is the best-measured diffractive process in small-$x$ DIS, with a body of $\rho$, $\phi$, and $J/\psi$ data from HERA \cite{ZEUS:2004yeh, H1:2005dtp, Ivanov:2004ax} and a well-developed theoretical description \cite{Ryskin:1992ui, Brodsky:1994kf, Collins:1996fb} which, in the dipole picture, takes the form of a convolution of the photon and meson light-cone wave functions with the dipole amplitude \cite{Nemchik:1994fp, Nemchik:1996cw, Munier:2001nr, Kowalski:2003hm, Forshaw:2003ki, Kowalski:2006hc}. Relative to dijets, the final state is cleaner since the final state does not rely on jet reconstruction. Furthermore, the polarization of the meson can be inferred from the angular distribution of the decay products (at least, in principle), which yields multiple observables for the same process (since the meson can be transversely or longitudinally polarized). On the other hand, the meson wave function is largely non-perturbative and must be modeled: this is indeed a drawback of this channel. (For a heavy quarkonium, one can instead use the non-relativistic expansion from \cite{Escobedo:2019bxn}, as was done in, e.g., \cite{Mantysaari:2022kdm}.) With this complication in mind, we show below that elastic vector meson production in polarized electron--proton collisions provides a complementary channel to elastic dijet production for measuring the OAM distribution since the same moment amplitudes appear in both channels.

In this paper we compute the longitudinal double-spin asymmetry (DSA) for elastic vector meson production in electron--proton collisions at small $x$ within the LCOT framework. We compute the DSA for both transversely and longitudinally polarized mesons. Our calculation focuses on the interference between longitudinally and transversely polarized virtual photons, which we refer to as the LT structure below. Expanding the LT structure in the DSA to the first (linear) order in $\Delta_\perp$, we find that the $\cos (\phi - \phi_\Delta)$ azimuthal harmonic, where $\phi$ and $\phi_{\Delta}$ are the azimuthal angles of the outgoing electron and vector meson, respectively, measured in the dipole/Trento frame \cite{Bacchetta:2004jz, Braun:2005rg, Mantysaari:2020lhf, Bhattacharya:2022vvo,Bhattacharya:2024sck,Hatta:2016aoc}, depends on a linear combination of the moment amplitudes $I_3$, $I_4$, $I_5$, and $I_E$ defined in \cite{Kovchegov:2023yzd, Becker:2026xcp} and below, weighted by known wave function overlaps. 
The expression also contains terms involving only the amplitudes already constrained by small-$x$ helicity phenomenology and includes the unpolarized dipole amplitude $N$, which is well known \cite{Balitsky:1995ub, Balitsky:1998ya, Kovchegov:1999yj, Kovchegov:1999ua}. The corresponding expressions for the virtual-photon cross section are given below in Eqs.~\eqref{lt_exp_cos} and \eqref{lt_exp_l_cos}, for transversly- and longitudinally-polarized vector mesons, respectively, and are the main results of this work.

The rest of this paper is organized as follows. Section~\ref{sec:setup} sets up the kinematics and decomposes the polarized cross sections into Lorentz-invariant structures. In Sec.~\ref{sec:production} we construct the elastic vector meson production amplitude at small $x$, working to sub-eikonal order and organizing the calculation by the type of polarized Wilson line operator involved. The double-spin asymmetry for transversely polarized mesons is computed in Sec.~\ref{sec:dsa_t} and for longitudinally polarized mesons in Sec.~\ref{sec:dsa_l}, with the expansion in $\Delta_\perp$ carried out in Secs.~\ref{sec:dsat_expansion} and \ref{sec:dsal_expansion} leading to the final results quoted above. We summarize and discuss the outlook for phenomenology at the future Electron-Ion Collider (EIC) \cite{Accardi:2012qut, Boer:2011fh, Proceedings:2020eah, AbdulKhalek:2021gbh} in Sec.~\ref{sec:conclusions}.

%%%%%%%%%%%%%%%%%%%%%%%%%%%%%%%%%%%%%%%%%%%%%%%

\section{General discussion of particle production in polarized electron-proton collisions}
\label{sec:setup}

We begin with a general discussion of particle production in polarized electron-proton scattering. In particular, consider a longitudinally polarized electron scattering off a longitudinally polarized proton. Using the notation of \cite{Diehl:2005pc}, we define the double- and single- spin asymmetries (DSA and SSA, respectively) as 
\begin{subequations}\label{A's}
\begin{align} \label{ALL}
 A_{LL}^{\gamma^*} &= \frac{ d\sigma (++) - d\sigma(+-) }{d\sigma (++) + d\sigma(+-)} \equiv \frac{ d\sigma (++) - d\sigma(+-) }{2 \, d\sigma_{\mathrm{unpol}}} \equiv   \frac{d \sigma^{\mathrm{DSA}} }{d\sigma_{\mathrm{unpol}}} , \\
 A_{UL}^{\gamma^*} &= \frac{ d\sigma (+) - d\sigma(-) }{d\sigma (+) + d\sigma(-)} = \frac{ d\sigma (+) - d\sigma(-) }{2 \, d\sigma_{\mathrm{unpol}}} \equiv   \frac{d \sigma^{\mathrm{SSA}} }{d\sigma_{\mathrm{unpol}}},
\end{align}
\end{subequations}
where $d\sigma(\sigma_eS_L)$ denotes the cross section for an electron and proton with helicities $\sigma_e$ and $S_L$, respectively. The numerators of the DSA and SSA along with the unpolarized cross section can therefore be written as
\begin{subequations} \label{asym_defs}
    \begin{align}
       \label{DSA_def}
 d \sigma^{\mathrm{DSA}} &= \frac{1}{4} \, \sum_{\sigma_e, S_L} \, \sigma_e \, S_L \, d \sigma (\sigma_e, S_L), \\
 \label{SSA_def}
 d \sigma^{\mathrm{SSA}}&= \frac{1}{4} \, \sum_{\sigma_e, S_L} \, S_L \, d \sigma (\sigma_e, S_L), \\
 d\sigma_{\mathrm{unpol}} &= \frac{1}{4} \, \sum_{\sigma_e, S_L} \, d \sigma (\sigma_e, S_L).
\end{align}
\end{subequations}
Below we will compute \eq{DSA_def} for elastic vector meson production. 

%%%%%%%%%%%%%%%%%%%%%%%%%%%%%%%%%%%%%%%
\begin{figure}[ht!]
    \centering
    \includegraphics[scale=0.5]{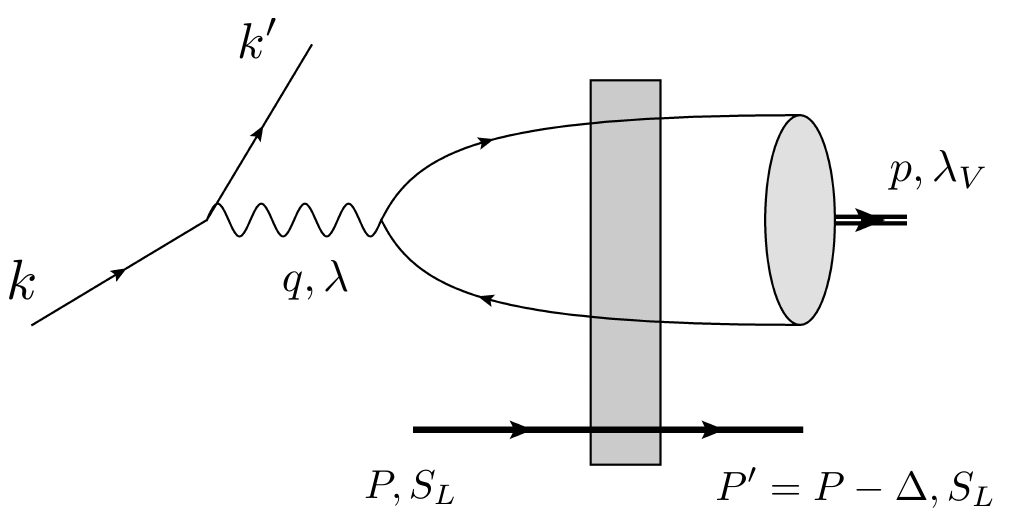}
    \caption{Leading diagram for elastic meson production at small $x$. The shaded vertical rectangle indicates the target shockwave, and the oval indicates the vector meson.}
    \label{fig:vmp}
\end{figure}
%%%%%%%%%%%%%%%%%%%%%%%%%%%%%%%%%%%%%%%

The leading diagram for particle production at small $x$ to lowest order in $\alpha_{EM}$ is shown in \fig{fig:vmp}. Following \cite{Bacchetta:2004jz, Braun:2005rg, Mantysaari:2020lhf, Bhattacharya:2022vvo, Bhattacharya:2024sck,Hatta:2016aoc, Kovchegov:2024wjs, Kovchegov:2024aus}, we adopt the dipole (or Trento) frame, in which the virtual photon carries no transverse momentum, $q_\perp = 0$, so that the relevant four-momenta take the form
\begin{subequations} \label{kinematics}
\begin{align}
P^\mu &\approx (P^+, 0^-, {\un 0}),  \\
 q^\mu &= \left( - \frac{Q^2}{2 q^-}, q^-, {\un 0} \right), \\
 k^\mu &= \left( \frac{k_\perp^2}{2 k^-} , k^-, {\un k} \right), \\
 k'^\mu &= \left( \frac{k_\perp^2}{2 (k^- - q^-)}, k^- - q^-, {\un k} \right).
\end{align}
\end{subequations}
We use the following definitions of the light-cone 4-vector components, $v^\pm = (v^0 \pm v^3)/\sqrt{2}$, while the transverse vectors are denoted by $\un v = (v^1, v^2)$ with their magnitude $v_\perp = |\un v|$.
As depicted in \fig{fig:vmp}, $P^\mu, q^\mu, k^\mu, k'^{\mu}$ denote the four-momenta of the incoming proton, virtual photon, incoming electron, and outgoing electron, respectively. The transverse momentum of the incoming electron is further decomposed as  $\un k = k_\perp (\cos \phi, \sin \phi)$. We also introduce the standard Lorentz invariants
\begin{subequations}\label{Lorentz_defs}
    \begin{align}
        Q^2 &= -q^2, \\ 
        y &= \frac{P \cdot q}{P \cdot k} = \frac{q^-}{k^-}, 
        \\ 
        x &= \frac{Q^2}{2 \, P\cdot q}, 
        \\ 
        s_{ep} &= (P + k)^2 \approx 2\,  P^+k^-, \\
        s & = (P+q)^2 \approx 2 P^+ q^- = y \, s_{ep}. \label{s_def}
    \end{align}
\end{subequations}
We note that in the $q_\perp =0$ Trento frame the (incoming) electron is not moving along the $z$ axis, and its and proton helicities measure spin projections on different directions. However, as shown in \cite{Diehl:2005pc}, one can replace the electron helicity by (twice) the electron spin projection on the $z$ axis in \eqs{A's}: at small $x$, the asymmetries would remain the same at the leading power of $x$.

In \cite{Kovchegov:2024wjs} (see also \cite{Diehl:2005pc, Diehl:1996st, Arens:1996xw, Kovchegov:2024aus}), it was shown that \eqs{asym_defs} can be written in terms of the virtual photon-proton cross section:
\begin{subequations} \label{observables}
    \begin{align} \label{DSA}
        E_{k^\prime} \frac{d \sigma^{\mathrm{DSA}}}{d^3 k^\prime} &= 
        \frac{\alpha_{EM}}{4\pi^2 \, Q^2}
         \half \, \sum_{S_L = \pm 1} S_L 
        \Bigg\{ 
            (2-y) \, \sum_{\lambda = \pm 1} \lambda \, \sigma^{\gamma^*p}_{\lambda \lambda} +  \sqrt{2(1-y)} \sum_{\lambda = \pm 1} 
            \Big[
            e^{i \lambda \phi} \sigma^{\gamma^*p}_{0 \lambda} + \mathrm{c.c.} 
            \Big] 
        \Bigg\},
        \\ 
        \notag 
    E_{k^\prime} \frac{d \sigma^{\mathrm{SSA}}}{d^3 k^\prime} &= 
    \frac{\alpha_{EM}}{4 \pi^2 \, Q^2 \, y} \, \half \, 
    \sum_{S_L = \pm 1} S_L 
        \Bigg\{ 
           \left[1+ (1-y)^2 \right] \sum_{\lambda = \pm 1} \sigma^{\gamma^*p}_{\lambda \lambda} - 2 (1-y) \Big[ e^{-2i\phi} \sigma^{\gamma^*p}_{1,-1} +\mathrm{c.c.}
        \Big]
        \\ 
        & \hspace{3cm}
          + 4 (1-y) \, \sigma^{\gamma^*p}_{00} +  (2-y) \sqrt{2(1-y)} \sum_{\lambda = \pm 1} \Big[
            \lambda \, e^{i \lambda \phi} \sigma^{\gamma^*p}_{0 \lambda} + \mathrm{c.c.}
          \Big]  
        \Bigg\},
        \label{SSA}
        \\ 
        \notag 
        E_{k^\prime} \frac{d \sigma_{\mathrm{unpol}}}{d^3 k^\prime} &= 
    \frac{\alpha_{EM}}{4 \pi^2 \, Q^2 \, y} \,\half \, 
    \sum_{S_L = \pm 1}
        \Bigg\{ 
           \left[1+ (1-y)^2 \right] \sum_{\lambda = \pm 1} \sigma^{\gamma^*p}_{\lambda \lambda} - 2 (1-y) \Big[ e^{-2i\phi} \sigma^{\gamma^*p}_{1,-1} +\mathrm{c.c.}
        \Big]
        \\ 
        & \hspace{3cm}
          + 4 (1-y) \, \sigma^{\gamma^*p}_{00} +  (2-y) \sqrt{2(1-y)} \sum_{\lambda = \pm 1} \Big[
            \lambda \, e^{i \lambda \phi} \sigma^{\gamma^*p}_{0 \lambda} + \mathrm{c.c.}
          \Big]  
        \Bigg\},
        \label{unpol}
    \end{align}
\end{subequations}
where $\sigma^{\gamma^* p}_{\lambda^\prime \lambda}$ denotes the virtual photon-proton cross section, with ($\lambda'$) $\lambda$ denoting the photon polarization in the (complex conjugate) amplitude. We denote longitudinal polarizations with $\lambda= 0$, while the transverse polarizations correspond to $\lambda = \pm 1$. Note that for $\lambda \neq \lambda'$, the object $\sigma^{\gamma^* p}_{\lambda^\prime \lambda}$ is not a cross section: rather, it is proportional to a component of the hadronic tensor in the photon helicity basis (see \cite{Kovchegov:2024wjs} and references therein). Above, $\alpha_{EM} = e^2/4 \pi$ is the fine structure constant with $e$ the electron charge.

\eq{DSA} is the main quantity we will calculate below. We will show that \eq{DSA} is a viable probe of the moment amplitudes, and therefore the OAM distributions. By contrast, for the elastic dijet case considered in \cite{Kovchegov:2024wjs}, we found that \eq{SSA} contained amplitudes with underdeveloped or non-existent small-$x$ phenomenology and therefore did not represent a viable observable for extracting the OAM distributions. For the elastic vector meson case at hand, we expect a similar statement to be true. In particular, the general argument presented in Sec.~VI of \cite{Kovchegov:2024wjs} is not restricted to elastic dijet production, and one can also use it to argue the SSA in elastic vector meson production would also not make a good probe for OAM. Therefore, we do not consider \eq{SSA} further in this paper, and we focus exclusively on the DSA in \eq{DSA}. Equation \eqref{unpol} 
can be evaluated using the eikonal small-$x$ techniques \cite{Mantysaari:2020lhf}, with all the contributions expressed in terms of the well-known unpolarized dipole amplitude $N$ \cite{Balitsky:1995ub, Balitsky:1998ya, Kovchegov:1999yj, Kovchegov:1999ua}. Once we compute \eq{DSA}, we may use \eqs{ALL} and (\ref{unpol}) to compute $A_{LL}^{\gamma^*}$.  

Here and throughout the rest of this paper, we consider the specific case of elastic vector meson production. To do so, introduce the four-momenta of the outgoing proton and vector meson (see \fig{fig:vmp})
\begin{subequations}
    \begin{align}
        P'^\mu &= \left( P^{\prime \, +}, \frac{P^{\prime \, 2}_T}{2 P^{\prime \, +}}, {\un P}' \right),
        \\ 
        p^\mu &= \left( \frac{\Delta_\perp^2 + M_V^2}{2\,p^-}, 
        p^- , 
        \un \Delta\right) 
    \end{align}
\end{subequations}
where $-\Delta^\mu$ is the momentum transfer to the proton and $M_V$ is the mass of the vector meson. In the dipole frame, ${\un P}'  = - \un \Delta$ . We further assume that $P^{\prime \, +}$ remains large (or, equivalently, $\Delta^+$ is small compared to $P^+$), so that $P' \cdot q \approx P^{\prime \, +} \, q^- \approx P^+ \, q^- \approx P \cdot q$. Corrections to this approximation are sub-eikonal, but do not couple to the proton longitudinal spin and therefore we neglect them here. 

Since the polarization of the vector meson can be discerned experimentally via the angular distribution of its decay products, we distinguish the numerator of the DSA separately for transverse and longitudinal mesons
\begin{subequations} \label{VMdsas}
\begin{align} \label{dsa_t}
    E_{k^\prime} \frac{d \sigma^{\mathrm{DSA}}_T}{d^3 k^\prime} &=
        \frac{\alpha_{EM}}{4\pi^2 \, Q^2}
         \half \, \sum_{S_L = \pm 1} S_L 
         \sum_{\lambda_V = \pm1}
        \Bigg\{ 
            (2-y) \, \sum_{\lambda = \pm 1} \lambda \, \sigma_{\lambda \lambda; \lambda_V}^{\gamma^*p \to Vp'} +  \sqrt{2(1-y)} \sum_{\lambda = \pm 1} 
            \Big[
            e^{i \lambda \phi} \sigma_{0 \lambda; \lambda_V}^{\gamma^*p \to V p'}+ \mathrm{c.c.} 
            \Big] 
        \Bigg\},
        \\  \label{dsa_l}
    E_{k^\prime} \frac{d \sigma^{\mathrm{DSA}}_L}{d^3 k^\prime} &=
        \frac{\alpha_{EM}}{4\pi^2 \, Q^2}
         \half \, \sum_{S_L = \pm 1} S_L 
        \Bigg\{ 
            (2-y) \, \sum_{\lambda = \pm 1} \lambda \, \sigma_{\lambda \lambda; \lambda_V=0}^{\gamma^*p \to Vp' } +  \sqrt{2(1-y)} \sum_{\lambda = \pm 1} 
            \Big[
            e^{i \lambda \phi} \sigma_{0 \lambda; \lambda_V= 0}^{\gamma^*p \to V p'}+ \mathrm{c.c.} 
            \Big] 
        \Bigg\},
\end{align}
\end{subequations}
where $\sigma^{\gamma^* p \to Vp'}_{\lambda' \lambda; \lambda_V}$ denotes the virtual photon-proton cross section for the elastic production of a vector meson $V$ with polarization $\lambda_V$. 
Each of \eqs{VMdsas} contains two structures. In the structure proportional to $(2-y)$, the virtual photon is transversely polarized in both the amplitude and the complex conjugate amplitude; we denote these terms the ``TT" channel. The other structures, proportional to $\sqrt{2(1-y)}$, arise from the interference between longitudinally and transversely polarized photons; we denote these the ``LT" channel. Only the latter can probe OAM, and in this work we compute only the LT structures. To see why, recall that the OAM distributions at small $x$ are determined by the moment amplitudes, which are first impact-parameter moments of the polarized dipole amplitudes \cite{Kovchegov:2023yzd, Becker:2026xcp}. They enter the cross section only once it is expanded to linear order in $\un \Delta$ (see Sections~\ref{sec:dsat_expansion} and \ref{sec:dsal_expansion} below as well as \cite{Kovchegov:2024wjs} for the elastic dijet case). In the dipole frame, after integration over the orientations of the dipoles, this single power of $\un \Delta$ has to be contracted with the only other transverse vector available, the electron momentum $\un k$ (forming either a dot-product or a cross-product). Any term linear in $\Delta_\perp$ must therefore depend on the electron azimuthal angle $\phi$. The TT structure carries no such dependence, as is evident from \eqs{VMdsas}, and hence contains no term linear in $\Delta_\perp$. For a transversely polarized meson in \eq{dsa_t}, it can be shown that the leading contribution to the TT term in the powers-of-$\Delta_\perp$ expansion is of order $(\Delta_\perp)^0$ and couples to the impact-parameter integrated polarized dipole amplitudes, which are related to the helicity PDFs at small $x$ but are insufficient to find the OAM distributions. For a longitudinally polarized meson, one can likewise show that the expansion of the TT term in \eq{dsa_l} in the powers of $\Delta_\perp$ starts at least at the order $\Delta_\perp^2$, where the cross section also involves second impact-parameter moments of the polarized dipole amplitudes, which are unrelated to OAM and currently theoretically and phenomenologically unknown. In neither case does the TT structure provide access to OAM, and we do not consider it further.

%%%%%%%%%%%%%%%%%%%%%%%%%%%%%%%%%%%%%%

\section{General calculation of the elastic vector meson production cross section to sub-eikonal order}

\label{sec:production}

In this Section, we compute the spin-dependent virtual photon-proton cross section, keeping the polarizations of the photon and the vector meson general. Then, in Sections~\ref{sec:dsa_t} and \ref{sec:dsa_l}, we compute the LT structures of \eqs{dsa_t} and (\ref{dsa_l}), respectively. 

In our infinite momentum frame, the virtual photon-proton cross section is 
\begin{align} \label{imf}
d\sigma^{\gamma^* p \to V p'} = \frac{1}{8 q^0 E_P} \, |M|^2 \, \frac{d^2 \Delta \, d \,p^-}{(2 \pi)^3 \, 2 p^-} \, \frac{d^2 P' \, d P^{\prime \, +}}{(2 \pi)^3 \, 2 P^{\prime \, +}} \, (2 \pi)^4 \, \delta^4 (q+P - p - P')   = |A|^2 \, \frac{d^2 \Delta}{(2 \pi)^2},
\end{align}
with $A = M/(2s)$, $M$ being the standard scattering amplitude. We thus arrive at 
\begin{align}\label{XS_VM1}
\half \,  \sum_{S_L = \pm 1} \, S_L \, \frac{d\sigma_{\lambda \lambda'; \lambda_V}^{\gamma^* p \to V p'}}{d^2 \Delta} = \frac{1}{4 \pi^2} \,  \half \,  \sum_{S_L = \pm 1} \, S_L \, A_{\lambda_V, \lambda'} \, A^*_{\lambda_V, \lambda} = \frac{1}{4 \pi^2} \,   \left[ A^\textrm{eik}_{\lambda_V, \lambda'} \, \left( A^\textrm{sub-eik}_{\lambda_V, \lambda} \right)^* +  A^\textrm{sub-eik}_{\lambda_V, \lambda'} \, \left( A^\textrm{eik}_{\lambda_V, \lambda} \right)^* \right].
\end{align}
To obtain the second equality in \eq{XS_VM1}, we have expanded the scattering amplitude in inverse powers of the center-of-mass energy squared to sub-eikonal order:
\begin{align} \label{eik-expansion}
A_{\lambda_V, \lambda} = A^\textrm{eik}_{\lambda_V, \lambda} + A^\textrm{sub-eik}_{\lambda_V, \lambda} + \ldots \, .
\end{align}
The eikonal term in \eq{eik-expansion} for our elastic vector meson process is well-known and reads (see \cite{Brodsky:1994kf,  Kovchegov:2012mbw} and references therein)
\begin{align} 
\notag
A^\textrm{eik}_{\lambda_V, \lambda} = i \, \int \frac{d^2 x_1 \, d^2 x_2}{4 \pi} \, \int\limits_0^1  \frac{dz}{z \, (1-z)} \, &e^{- i \un \Delta \cdot \left[ z \, \un x_1 + (1-z) \, \un x_2 \right]} 
\\  & 
\times\sum_{\sigma_1, \sigma_2} \, \Psi_{\lambda, \sigma_1, \sigma_2}^{\gamma^* \to q {\bar q}} ({\un x}_{12}, z) 
 \left\langle \tord \tr \left[ 1 - V_{\un 1} \, V_{\un 2}^\dagger \right] \right\rangle \left[ \Psi_{\lambda_V, \sigma_1, \sigma_2}^{V} ({\un x}_{12}, z) \right]^*
, 
\label{A_eik}
\end{align}
where the angle brackets denote the standard saturation physics averaging in the (unpolarized) proton wavefunction, $\Psi_{\lambda, \sigma_1, \sigma_2}^{\gamma^* \to q {\bar q}}$ denotes the light cone wave function for the virtual photon splitting into a $q \bar q$ pair, and $\Psi_{\lambda_V, \sigma_1, \sigma_2}^{V}$ is the vector meson light-cone wave function \cite{Lepage:1980fj, Brodsky:1997de}. Both wavefunctions are taken in a mixed representation: they are in the transverse position space, and in the longitudinal momentum space. We label the transverse position of the quark from the $q \bar q$ pair by $\un x_1$, while the anti-quark is located at $\un x_2$: their separation is $\un x_{12} = \un x_1 - \un x_2$. In addition, $z$ is the fraction of the $q^-$ momentum of the virtual photon carried by the quark in the  $q \bar q$ pair (see \fig{fig:diagrams} below).

We define the eikonal fundamental Wilson line as 
\begin{align}\label{Wilson_line}
   V_{\un x}[x_f^-, x_i^-] = \mathcal{P} \exp \left[ig \int \displaylimits^{x_f^-}_{x_i^-} dx^- A^+(0^+, x^-, \un x)  \right],
\end{align}
where $A^+ = t^a A^{+a}$ is the gluon field of the target with $t^a$ the generators of SU($N_c$) in the fundamental representation and $g$ is the strong coupling constant. We abbreviate infinite light-cone Wilson lines as $V_{\un x}  \equiv V_{\un x}[\infty, -\infty]$. For transverse positions $\un x_i$, we will abbreviate $V_{\un x_i} = V_{\un i}$. Defining the (unpolarized) dipole $S$-matrix \cite{Mueller:1994rr,Mueller:1994jq,Mueller:1995gb,Balitsky:1995ub,Balitsky:1998ya,Kovchegov:1999yj,Kovchegov:1999ua,Jalilian-Marian:1997dw,Jalilian-Marian:1997gr,Weigert:2000gi,Iancu:2001ad,Iancu:2000hn,Ferreiro:2001qy},
\begin{align} \label{unpol-dip-amp}
S_{10} (s) =  \frac{1}{N_c} \, \left\langle \mbox{T} \, \tr \left[ V_{\un 1} \,  V_{\un 0}^{\dagger} \right] \right\rangle (s),
\end{align}
we may write \eq{A_eik} as
\begin{align} 
\notag
A^\textrm{eik}_{\lambda_V, \lambda} = i \, N_c\int \frac{d^2 x_1 \, d^2 x_2}{4 \pi} \, \int\limits_0^1  \frac{dz}{z \, (1-z)} \, &e^{- i \un \Delta \cdot \left[ z \, \un x_1 + (1-z) \, \un x_2 \right]} 
\\  & 
\times\sum_{\sigma_1, \sigma_2} \, \Psi_{\lambda, \sigma_1, \sigma_2}^{\gamma^* \to q {\bar q}} ({\un x}_{12}, z) 
\Big[  1- S_{12}(s)\Big] \left[ \Psi_{\lambda_V, \sigma_1, \sigma_2}^{V} ({\un x}_{12}, z) \right]^*.
\label{A_eik2}
\end{align}
It is convenient to decompose the unpolarized dipole $S$-matrix into the $\mathcal{C}$-even Pomeron amplitude $N_{10}(s)$ and the $\mathcal{C}$-odd odderon amplitude $O_{10}(s)$ \cite{Kovchegov:2003dm,Hatta:2005as},
\begin{align} \label{unpdas}
    1- S_{10}(s) &= N_{10}(s) - i O_{10}(s),
\end{align}
with 
\begin{subequations}
\begin{align}
& N_{10} (s) = 1 - \frac{1}{2 N_c} \, \left\langle \tord \tr \left[ V_{\un 1} \, V^\dagger_{\un 0} \right] + \tord \tr \left[ V_{\un 0} \, V^\dagger_{\un 1} \right] \right\rangle(s), \\
& O_{10} (s) = \frac{1}{2 N_c \, i} \, \left\langle \tord \tr \left[ V_{\un 1} \, V^\dagger_{\un 0} \right] - \tord \tr \left[ V_{\un 0} \, V^\dagger_{\un 1} \right] \right\rangle(s).
\end{align}
\end{subequations}
Both amplitudes have been studied extensively (see \cite{Mueller:1994rr,Mueller:1994jq,Mueller:1995gb,Balitsky:1995ub,Balitsky:1998ya,Kovchegov:1999yj,Kovchegov:1999ua,Jalilian-Marian:1997dw,Jalilian-Marian:1997gr,Weigert:2000gi,Iancu:2001ad,Iancu:2000hn,Ferreiro:2001qy} for the Pomeron amplitude and  \cite{Kovchegov:2003dm,Hatta:2005as} for the odderon amplitude). In the single-logarithmic approximation (SLA), which resums powers of $\as \ln (1/x)$, and for the quasi-classical initial conditions \cite{Mueller:1989st, McLerran:1993ka, McLerran:1993ni, McLerran:1994vd},
both amplitudes are real, and we assume this throughout. The odderon, however, does not contribute to \eq{A_eik2} at all. For vector meson wave functions symmetric under $z \to 1-z$, the overlaps $\sum_{\sigma_1, \sigma_2} \Psi_{\lambda, \sigma_1, \sigma_2}^{\gamma^* \to q {\bar q}} ({\un x}_{12}, z) \left[ \Psi_{\lambda_V, \sigma_1, \sigma_2}^{V} ({\un x}_{12}, z) \right]^*$, given by \eqs{type2_overlaps} below (for the case where both transverse separations in \eqs{type2_overlaps} are the same $\un x = \un x_{12}$), are invariant under the simultaneous interchange $\un x_1 \leftrightarrow \un x_2$, $z \leftrightarrow 1-z$ for all $\lambda$ and $\lambda_V$. The Fourier phase is invariant as well, whereas $O_{12}(s) = - O_{21}(s)$, so the odderon term integrates to zero. This is a manifestation of $\mathcal{C}$-parity. The virtual photon and the vector meson are both $\mathcal{C}$-odd, so the $\gamma^* \to V$ transition requires a $\mathcal{C}$-even exchange in the $t$-channel. (Note that the odderon should contribute to the production of positive $\mathcal{C}$-parity mesons, such as the pseudoscalar mesons \cite{Ewerz:2003xi}.) The eikonal amplitude therefore reduces to
\begin{align} 
A^\textrm{eik}_{\lambda_V, \lambda} = i \, N_c\int \frac{d^2 x_1 \, d^2 x_2}{4 \pi} \, \int\limits_0^1  \frac{dz}{z \, (1-z)} \, e^{- i \un \Delta \cdot \left[ z \, \un x_1 + (1-z) \, \un x_2 \right]} 
\sum_{\sigma_1, \sigma_2} \, \Psi_{\lambda, \sigma_1, \sigma_2}^{\gamma^* \to q {\bar q}} ({\un x}_{12}, z) \, 
N_{12}(s) \left[ \Psi_{\lambda_V, \sigma_1, \sigma_2}^{V} ({\un x}_{12}, z) \right]^*.
\label{A_eik3}
\end{align}
Here we have also employed the fact, substantiated by the $z$-dependence of the light-cone wavefunctions, that the $z$-integral generates no large logarithms coming from the $z \to 0$ and $z \to 0$ regions: hence, the argument of the dipole amplitude $N_{12}$ is $s$, and not, for instance, $zs$. 

The light-cone wavefunctions for a photon to split into a quark-antiquark pair are known \cite{Kovchegov:2012mbw}
\begin{subequations}\label{LCwf}
\begin{align} 
 \notag
& \Psi_{\lambda = \pm 1, \sigma_1, \sigma_2}^{\gamma^* \to q {\bar q}} ({\un x}_{12}, z) = \frac{e Z_f}{2 \pi} \, \sqrt{z \, (1-z)} \,  \Bigg[ \delta_{\sigma_1, - \sigma_2} \, (1-2 z + \sigma_1 \, \lambda) \, i a_f \, \frac{{\un \epsilon}_\lambda \cdot {\un x}_{12}}{x_{12}} \, K_1 \left( x_{12} \, a_f \right) \\ 
& \hspace*{8cm} + \delta_{\sigma_1 \sigma_2} \, \frac{m_f}{\sqrt{2}} \, (1+ \sigma_1 \, \lambda) \, K_0 \left( x_{12} \, a_f \right) \Bigg] , \\ \label{long_lcwf}
& \Psi_{\lambda = 0, \sigma_1, \sigma_2}^{\gamma^* \to q {\bar q}} ({\un x}_{12}, z) = - \frac{e Z_f}{2 \pi} \, \delta_{\sigma_1, - \sigma_2} \,  [z \, (1-z)]^{3/2} \, 2 Q \, K_0 \left( x_{12} \, a_f \right), 
\end{align}
\end{subequations}
where $a_f^2 = Q^2z(1-z) + m_f^2$, with $m_f$ being the mass of the quark of flavor $f$. The virtual photon carries either the transverse (T) or longitudinal (L) polarizations, which, for $q_\perp =0$ are
\begin{align}
\epsilon_T^\lambda = (0,0, {\un \epsilon}_\lambda ) , \ \ \ \epsilon_L = \left( \frac{Q}{2 q^-}, \frac{q^-}{Q}, {\un 0} \right)
\end{align}
with ${\un \epsilon}_\lambda = - (1/\sqrt{2}) (-\lambda, i)$. $Z_f$ denotes the fractional electric charge of the quark of flavor $f$, while the quark and anti-quark helicities are labeled $\sigma_1$ and $\sigma_2$, respectively.

Note that \eq{long_lcwf} also contains a term $\sim \delta^2 (\un x_{10})$ (not shown above), which is zero at the eikonal order when multiplying a dipole amplitude. It appears to be zero at the sub-eikonal order as well, but only when multiplying a dipole amplitude (and not, say, a quadrupole). The same applies to the $\gamma^* \to q {\bar q}$ splitting happening inside the shock wave \cite{Kovchegov:2025gcg}: it results in $\delta^2 (\un x_{10})$ which is zero when multiplying a dipole amplitude. Since we only have dipole amplitudes in this problem, we drop all such $\sim \delta^2 (\un x_{10})$ terms. 

For the vector meson wavefunction, we follow \cite{Kowalski:2003hm, Nemchik:1994fp, Nemchik:1996cw, Dosch:1996ss, Forshaw:2003ki, Kowalski:2006hc} (see also \cite{Kovchegov:2023bvy}), and assume that the vector meson is predominantly a quark--antiquark state that has the same polarization structure as a virtual photon. Via the replacement
\begin{equation}
    \frac{eZ_{f}}{2 \pi} z(1-z) K_{0}(x_{\perp}a_{f}) \rightarrow \phi_{T, L}(x_{\perp},z),
\end{equation}
of the scalar part of the wave function for the longitudinally (L) or transversely (T) polarized vector meson, we can write down the vector meson wave functions as
\begin{subequations}
\begin{align}
&  \Psi_{\lambda_V = \pm 1, \sigma_1, \sigma_2}^{V}({\un x}_{12},z)=\frac{1}{\sqrt{z(1-z)}} \left[ -i \, \delta_{\sigma_1, - \sigma_2} \, (1-2z + \sigma_1 \, \lambda_V)\frac{{\un \epsilon}_{\lambda_V} \cdot {\un x}_{12}}{x_{12}} \, \frac{\pd}{\pd x_{12}} +\delta_{\sigma_1 \sigma_2}\frac{m_f}{\sqrt{2}}(1+\sigma_1 \, \lambda_V)\right]\phi_{T} (x_{12},z), \\
& \Psi_{\lambda_V = 0, \sigma_1, \sigma_2}^{V} ({\un x}_{12}, z) = - \delta_{\sigma_1, - \sigma_2} \,  \sqrt{z \, (1-z)} \, \left[ M_V + \delta \, \frac{m_f^2 - \nabla_{12}^2}{ M_V \, z \, (1-z)} \right] \, \phi_{L}(x_{12},z) ,
\end{align}
\end{subequations}
where we have also replaced \cite{Nemchik:1994fp, Nemchik:1996cw}
\begin{align}
2 Q \to M_V + \delta \, \frac{m_f^2 - \nabla_{12}^2}{ M_V \, z \, (1-z)} 
\end{align}
in the longitudinal vector meson wave function with a dimensionless parameter $\delta =0, 1$. The functions $\phi_{T} (x_{12},z)$ and $\phi_{L}(x_{12},z)$ depend on the model of the vector meson: some frequently used examples can be found in \cite{Kowalski:2003hm, Nemchik:1994fp, Nemchik:1996cw, Dosch:1996ss, Forshaw:2003ki, Kowalski:2006hc, Kovchegov:2023bvy}.

To compute the sub-eikonal term in \eq{eik-expansion}, we employ the LCOT formalism \cite{Kovchegov:2015pbl, Hatta:2016aoc, Kovchegov:2016zex, Kovchegov:2016weo, Kovchegov:2017jxc, Kovchegov:2017lsr, Kovchegov:2018znm, Kovchegov:2019rrz, Boussarie:2019icw, Cougoulic:2019aja, Kovchegov:2020hgb, Cougoulic:2020tbc, Chirilli:2021lif, Kovchegov:2021lvz, Cougoulic:2022gbk, Borden:2023ugd, Adamiak:2023okq, Borden:2024bxa}. In the LCOT formalism, one expands the quark-proton $S$-matrix to sub-eikonal order:
\begin{align} \label{Wilsonl}
    V_{\un x, \un y; \sigma^\prime, \sigma} = V_{\un x} \, \delta^2(\un x - \un y) \, \delta_{\sigma, \sigma^\prime} + V^{\mathrm{pol}}_{\un x, \un y; \sigma^\prime, \sigma},
\end{align}
where $\un x$ ($\un y$) denotes the transverse positions of the incoming (outgoing) quark and $\sigma\, (\sigma')$ denotes the helicity of the incoming (outgoing) quark. It has been shown \cite{Kovchegov:2017lsr, Kovchegov:2018znm, Kovchegov:2018zeq, Chirilli:2018kkw, Altinoluk:2020oyd, Kovchegov:2021iyc, Cougoulic:2022gbk} that the sub-eikonal piece of \eq{Wilsonl} can be written as
\begin{align} \label{q-Smatrix}
    V_{\un x, \un y; \sigma^\prime, \sigma}^{\mathrm{pol}} = \sigma \delta_{\sigma, \sigma^\prime} \Big[V^{\mathrm{G}[1]}_{\un x} + V^{\mathrm{q}[1]}_{\un x}  \Big] \delta^2(\un x - \un y)
    + \delta_{\sigma, \sigma^\prime} \Big[V^{\mathrm{G}[2]}_{\un x, \un y} + V^{\mathrm{q}[2]}_{\un x} \delta^2(\un x - \un y)  \Big],
\end{align}
where we have defined the polarized Wilson lines as \cite{Cougoulic:2022gbk}
\begin{subequations}\label{VqG}
\begin{align}
& V_{\un x}^{\textrm{G} [1]}  = \frac{i \, g \, P^+}{s} \int\limits_{-\infty}^{\infty} d{x}^- V_{\un{x}} [ \infty, x^-] \, F^{12} (x^-, {\un x}) \, \, V_{\un{x}} [ x^-, -\infty]  , \label{VG1} \\
& V_{\un x}^{\textrm{q} [1]}  = \frac{g^2 P^+}{2 \, s} \int\limits_{-\infty}^{\infty} \!\! d{x}_1^- \! \int\limits_{x_1^-}^\infty d x_2^- V_{\un{x}} [ \infty, x_2^-] \, t^b \, \psi_{\beta} (x_2^-,\un{x}) \, U_{\un{x}}^{ba} [x_2^-, x_1^-] \, \left[ \gamma^+ \gamma^5 \right]_{\alpha \beta} \, \bar{\psi}_\alpha (x_1^-,\un{x}) \, t^a \, V_{\un{x}} [ x_1^-, -\infty] , \label{Vq1} \\
& V_{{\ul x}, {\un y}}^{\textrm{G} [2]}  = - \frac{i \, P^+}{s} \int\limits_{-\infty}^{\infty} d{z}^- d^2 z \ V_{\un{x}} [ \infty, z^-] \, \delta^2 (\un{x} - \un{z}) \, \cev{D}^i (z^-, {\un z}) \, D^i  (z^-, {\un z}) \, V_{\un{y}} [ z^-, -\infty] \, \delta^2 (\un{y} - \un{z}) , \label{VxyG2} \\
& V_{{\ul x}}^{\textrm{q} [2]} = - \frac{g^2 P^+}{2 \, s} \int\limits_{-\infty}^{\infty} \!\! d{x}_1^- \! \int\limits_{x_1^-}^\infty d x_2^- V_{\un{x}} [ \infty, x_2^-] \, t^b \, \psi_{\beta} (x_2^-,\un{x}) \, U_{\un{x}}^{ba} [x_2^-, x_1^-] \, \left[ \gamma^+ \right]_{\alpha \beta} \, \bar{\psi}_\alpha (x_1^-,\un{x}) \, t^a \, V_{\un{x}} [ x_1^-, -\infty] \label{Vq2},
\end{align}
\end{subequations}
where $F^{12}$ is the gluon field strength tensor, and $s$ is the invariant mass of the quark-proton system. The right- and left-acting covariant derivatives are defined as $D^i = \partial^i - i g A^i$ and $\cev{D}^i = \cev{\partial}^i + i g A^i$, respectively. Additionally, the adjoint eikonal Wilson line $U_{\un x}[x_f^-, x_i^-]$ is obtained by replacing $A^+$ in \eq{Wilson_line} with $\mathcal{A}^+ = \mathcal{A}^{+a} T^a$, the target gluon field in the adjoint representation. 

In addition to the operators in \eqs{VqG}, there is another operator at sub-eikonal order \cite{Chirilli:2018kkw, Chirilli:2021lif, Altinoluk:2021lvu, Li:2023tlw},
\begin{align}
V_{\un x}^{\textrm{G} [3]}  = \frac{i \, g \, P^+}{s} \int\limits_{-\infty}^{\infty} d{x}^- V_{\un{x}} [ \infty, x^-] \, F^{+-} (x^-, {\un x}) \, \, V_{\un{x}} [ x^-, -\infty] . \label{VG3} 
\end{align}
\eq{VG3} can be understood by replacing $0^+$ in \eq{Wilson_line} by $x^+$ and expanding in $x^+$ (see \cite{Kovchegov:2025gcg, Chirilli:2018kkw, Chirilli:2021lif, Altinoluk:2021lvu, Altinoluk:2023lly, Li:2023tlw, Mukherjee:2026cte}). For instance, in $A^-= 0$ gauge, we write
\begin{align} \notag
    \int\limits_{-\infty}^\infty d x^+ \, e^{i (p_f^- - p_i^-) \, x^+} \, V_{\un{x}} (x^+) &= \int\limits_{-\infty}^\infty d x^+ \, e^{i (p_f^- - p_i^-) \, x^+} \, \left[ V_{\un{x}} (0^+) + x^+ \, \pd^- V_{\un{x}} (0^+) + \ldots \right] \\
    & = 2 \pi \, \delta (p_f^- - p_i^-) \, V_{\un{x}} (0^+) +  2 \pi \, i \, \left[ \frac{\pd}{\pd p_f^-} \delta (p_f^- - p_i^-) \right] \, 2 \sqrt{p_f^- \, p_i^-} \, V_{\un x}^{\textrm{G} [3]} + \ldots \, . 
    \label{V_expansion}
\end{align}
By contrast, the operators in \eqs{VqG} enter our expressions for the vector meson amplitude below with the same minus-momentum conserving delta function $2 \pi \, \delta(p_f^- - p_i^-)$ as the eikonal term (the first term in the second line of \eq{V_expansion}).

When discussing the sub-eikonal operators in \eqs{VqG} and \eqref{VG3} above, we have, for simplicity of notation, momentarily assumed that $s$ is the quark--proton center-of-mass energy squared. Now, we return to its definition in \eq{s_def} as the virtual photon--proton center of mass energy squared.

For the type-1 and type-2 operators, the sub-eikonal amplitude for our elastic vector meson process can then be written as
\begin{align}
 \notag 
A^\textrm{sub-eik}_{\lambda_V, \lambda} &=  - i \, \int \frac{d^2 x_1 \, d^2 x_0 \, d^2 x_2}{4 \pi} \, \int\limits_0^1 \frac{dz}{z \, (1-z)} \, e^{- i \un \Delta \cdot \left[ z \, \un x_1 + (1-z) \, \un x_2 \right]} \,  \\ 
 &\times \, \sum_{\sigma_1, \sigma_2, \sigma'} \, \Bigg \{ \Psi_{\lambda, \sigma', \sigma_2}^{\gamma^* \to q {\bar q}} ({\un x}_{02}, z) \, \left\langle \tord \tr \left[ V^\textrm{pol}_{\un 1, \un 0; \sigma_1, \sigma'} \, V_{\un 2}^\dagger \right] \right\rangle (zs) \, \left[ \Psi_{\lambda_V, \sigma_1, \sigma_2}^{V} ({\un x}_{12}, z) \right]^* \notag \\
&  \hspace{2cm}
- \Psi_{\lambda, \sigma_1, \sigma'}^{\gamma^* \to q {\bar q}} ({\un x}_{10}, z) \, \left\langle \tord \tr \left[ V_{\un 1} \, V^{\textrm{pol} \, \dagger}_{\un 2, \un 0; -\sigma_2, - \sigma'}  \right] \right\rangle ((1-z) s) \, \left[ \Psi_{\lambda_V, \sigma_1, \sigma_2}^{V} ({\un x}_{12}, z) \right]^* \Bigg\},
\label{A_subeik}
\end{align}
where we have used the fact the antiquark $S$-matrix is related to the quark one via $\overline{V}_{\un x, \un y; \sigma, \sigma'} = -\left(V_{\un x, \un y; -\sigma , -\sigma'} \right)^\dagger$ \cite{Borden:2024bxa, Kovchegov:2024wjs, Kovchegov:2025gcg}. \eq{A_subeik} is illustrated in \fig{fig:diagrams}. Note that $\un x_{ij} = \un x_i - \un x_j$. The type-3 contributions to the elastic vector meson amplitude are treated in Sec.~\ref{sec:type-3}. 

%%%%%%%%%%%%%%%%%%%%%%%%%%%%%%%%%%%%%%%%%%%%%%%%%%%%%%%%%%%%%%%%%%%%%%%%%%%%%%%%%%%%
\begin{figure}[ht!]
\centering
\includegraphics[scale=0.45]{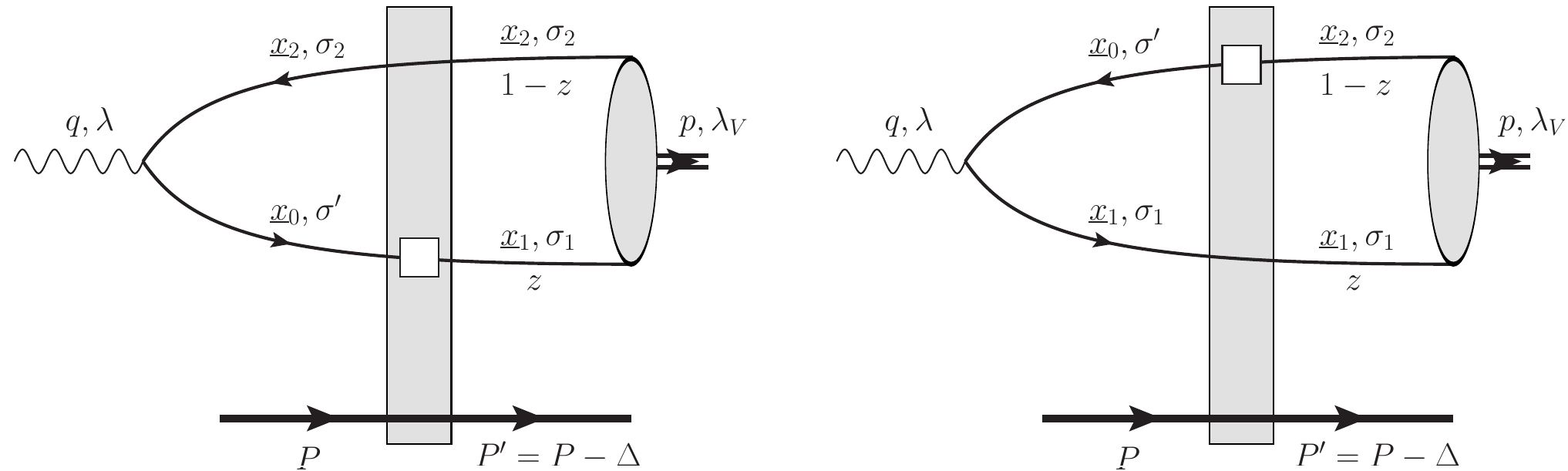}  
\caption{Elastic vector meson production diagrams at the sub-eikonal level. The vertical shaded rectangles represent the proton shockwave and the square boxes indicate the sub-eikonal parts of the quark and antiquark $S$-matrices. The shaded oval indicates the vector meson light-cone wavefunction.}
\label{fig:diagrams}
\end{figure}
%%%%%%%%%%%%%%%%%%%%%%%%%%%%%%%%%%%%%%%%%%%%%%%%%%%%%%%%%%%%%%%%%%%%%%%%%%%%%%%%%%%%

It is important to clarify here that the vector meson wave function often cuts off the $z$-values close to $0$ and $1$ by some energy-independent cutoff. For instance, in the ``boosted Gaussian" model \cite{Nemchik:1994fp, Nemchik:1996cw, Dosch:1996ss, Forshaw:2003ki, Kowalski:2003hm, Kowalski:2006hc} one has
\begin{align}\label{bG}
\phi_{L,T} (x_{12},z) \sim  z (1-z) \, e^{- \frac{m_f^2 \, R^2}{8 z (1-z)}}    
\end{align}
with some meson radius parameter $R$. This dependence makes sure that $z > m_f^2 \, R^2$ and $1-z > m_f^2 \, R^2$. At the same time, in the shockwave picture we employ, the requirement that the dipole lifetime is longer than the $x^-$-width of the shockwave leads to $z > 1/(s \, x_{12}^2) > \Lambda^2/s$ and $1-z > 1/(s \, x_{12}^2) > \Lambda^2/s$, which is implied in all our $z$-integrals (with $\Lambda$ an infrared cutoff). This way, further evaluation of \eq{A_subeik} performed below retains the logarithmic divergences at $z \to 0$ and $z \to 1$ in the $z$-integral, regardless of whether these divergences are cut off by $m_f^2 \, R^2$ or by $\Lambda^2/s$ (that is, regardless of which one of these two parameters is larger) .

%%%%%%%%%%%%%%%%%%%%%%%%%%%%%%%%%%%%%%%%%

\subsection{Type-1 operators}

The contribution of the type-1 operators to \eq{A_subeik} reads 
\begin{align}
 \notag 
A^\textrm{[1] sub-eik}_{\lambda_V, \lambda} &=  - \frac{i}{s} \, \int \frac{d^2 x_1 \, d^2 x_2}{4 \pi} \, \int\limits_0^1 \frac{dz}{z \, (1-z)} \, e^{- i \un \Delta \cdot \left[ z \, \un x_1 + (1-z) \, \un x_2 \right]} \,  \sum_{\sigma_1, \sigma_2} \, \Psi_{\lambda, \sigma_1, \sigma_2}^{\gamma^* \to q {\bar q}} ({\un x}_{12}, z) \, \left[ \Psi_{\lambda_V, \sigma_1, \sigma_2}^{V} ({\un x}_{12}, z) \right]^* 
\\ 
& \times \, \left[ \frac{\sigma_1}{z} \, \llangle \tord \tr \left[ V^{\textrm{pol} [1]}_{\un 1} \, V_{\un 2}^\dagger \right] \rrangle (z s) + \frac{\sigma_2}{1-z} \, \llangle \tord \tr \left[ V_{\un 1} \, V^{\textrm{pol} [1] \, \dagger}_{\un 2}  \right] \rrangle ((1-z) s) \right] ,
\label{A_subeik_11}
\end{align}
where we have defined $V^{\mathrm{pol}[1]}_{\un x} = V^{\mathrm{G}[1]}_{\un x} + V^{\mathrm{q}[1]}_{\un x}$ and employed double bracket notation, which is defined as \cite{Kovchegov:2015pbl, Kovchegov:2017lsr}
\begin{align} \label{double-bracket}
    \llangle \ldots \rrangle(s) = s\,  \Big\langle \ldots \Big\rangle(s).
\end{align}

Additionally, for simplicity, we define the wavefunction overlaps in \eq{A_subeik_11} for the various photon polarization channels as 
\begin{subequations} \label{type1_overlaps}
\begin{align}
 \notag 
& \textrm{\bf TT:} \hspace{0.5cm} \sum_{\sigma_1, \sigma_2} \, \sigma_{p} \,  \Psi_{\lambda = T, \sigma_1, \sigma_2}^{\gamma^* \to q {\bar q}} ({\un x}_{12}, z) \, \left[ \Psi_{\lambda_V = T, \sigma_1, \sigma_2}^{V} ({\un x}_{12}, z) \right]^* 
\\ \notag & \hspace{3cm}
= \frac{e Z_f}{\pi} \, \lambda \, \delta_{\lambda \lambda_V} \, \Bigg[ (-1)^p \, (1-2z) \, a_f \, K_1 (x_{12} \, a_f) \, \frac{\pd \phi_{T} (x_{12},z)}{\pd x_{12}}  + m_f^2 \, K_0 (x_{12} \, a_f) \, \phi_{T} (x_{12},z) \Bigg] 
\\ & \hspace{3cm}
\equiv \lambda \, \delta_{\lambda \lambda_V} \, \left[ (-1)^p \, \Phi^{[1]}_{TT} (x_{12},z) + \Phi^{\prime \, [1]}_{TT} (x_{12},z) \right] ,\\
 \notag
& \textrm{\bf TL:} \hspace{0.5cm} \sum_{\sigma_1, \sigma_2} \, \sigma_{p} \,  \Psi_{\lambda = T, \sigma_1, \sigma_2}^{\gamma^* \to q {\bar q}} ({\un x}_{12}, z) \, \left[ \Psi_{\lambda_V = 0, \sigma_1, \sigma_2}^{V} ({\un x}_{12}, z) \right]^* 
\\ \notag & \hspace{3cm}
= (-1)^p \, \frac{e Z_f}{\pi} \, z (1-z) \, i \, a_f \, \lambda \, \frac{{\un \epsilon}_{\lambda} \cdot {\un x}_{12}}{x_{12}} \, K_1 (x_{12} \, a_f)  \left[ M_V + \delta \, \frac{m_f^2 - \nabla_{12}^2}{ M_V \, z \, (1-z)} \right] \, \phi_{L}(x_{12},z)
\\  & \hspace{3cm}
\equiv i \, \lambda \, \frac{{\un \epsilon}_{\lambda} \cdot {\un x}_{12}}{x_{12}} \, (-1)^p \, \Phi^{[1]}_{TL} (x_{12},z ) , \\
\notag
& \textrm{\bf LT:} \hspace{0.5cm} \sum_{\sigma_1, \sigma_2} \, \sigma_{p} \,  \Psi_{\lambda = 0, \sigma_1, \sigma_2}^{\gamma^* \to q {\bar q}} ({\un x}_{12}, z) \, \left[ \Psi_{\lambda_V = T, \sigma_1, \sigma_2}^{V} ({\un x}_{12}, z) \right]^* 
\\ \notag & \hspace{3cm}
= (-1)^p \, \frac{e Z_f}{\pi} \, z (1-z) \, i \, 2 Q \, K_0 (x_{12} \, a_f) \, \lambda_V \, \frac{{\un \epsilon}^*_{\lambda_V} \cdot {\un x}_{12}}{x_{12}} \, \frac{\pd \phi_{T} (x_{12},z)}{\pd x_{12}} 
\\  & \hspace{3cm}
\equiv i \, \lambda_V \, \frac{{\un \epsilon}^*_{\lambda_V} \cdot {\un x}_{12}}{x_{12}} \, (-1)^p \, \Phi^{[1]}_{LT} (x_{12},z) ,  \\
& \textrm{\bf LL:} \hspace{0.5cm} \sum_{\sigma_1, \sigma_2} \, \sigma_{p} \,  \Psi_{\lambda = 0, \sigma_1, \sigma_2}^{\gamma^* \to q {\bar q}} ({\un x}_{12}, z) \, \left[ \Psi_{\lambda_V = 0, \sigma_1, \sigma_2}^{V} ({\un x}_{12}, z) \right]^* = 0. \label{LL_type-1}
\end{align}
\end{subequations}
Here $p=1,2$. Explicit expressions for the $\Phi$ functions we will use below are given in Appendix~\ref{app:overlaps}. Note all the $\Phi$ functions appearing in \eqs{type1_overlaps} and below in \eqs{type2_overlaps} and \eqref{type3_overlaps} are real. 

%%%%%%%%%%%%%%%%%%%%%%%%%%%%%%%%%%%%%%%%%

\subsection{Type-2 operators}

Moving on to the contribution of the type-2 operators to \eq{A_subeik}, we get for the contribution of $V_{\un x}^{\textrm{q} [2]}$
\begin{align}
 \notag 
 A^\textrm{q[2] sub-eik}_{\lambda_V, \lambda} &=  - \frac{i}{s} \, \int \frac{d^2 x_1 \, d^2 x_2}{4 \pi} \, \int\limits_0^1 \frac{dz}{z \, (1-z)} \, e^{- i \un \Delta \cdot \left[ z \, \un x_1 + (1-z) \, \un x_2\right]} \,  \sum_{\sigma_1, \sigma_2} \, \Psi_{\lambda, \sigma_1, \sigma_2}^{\gamma^* \to q {\bar q}} ({\un x}_{12}, z) \, \left[ \Psi_{\lambda_V, \sigma_1, \sigma_2}^{V} ({\un x}_{12}, z) \right]^* 
\\ 
& \hspace{3cm}
\times \, \left[ \frac{1}{z} \, \llangle \tord \tr \left[ V^{\textrm{q} [2]}_{\un 1} \, V_{\un 2}^\dagger \right] \rrangle (z s) - \frac{1}{1-z} \, \llangle \tord \tr \left[ V_{\un 1} \, V^{\textrm{q} [2] \, \dagger}_{\un 2}  \right] \rrangle ((1-z) s) \right] .
\label{A_subeik_2q}
\end{align}
Similarly, the contribution of $V_{{\un x}, {\un y}}^{\textrm{G} [2]}$ is
\begin{align}
\notag 
 A^\textrm{G[2] sub-eik}_{\lambda_V, \lambda} &=  - \frac{i}{s} \, \int \frac{d^2 x_1 \, d^2 x_0 \, d^2 x_2}{4 \pi} \, \int\limits_0^1 \frac{dz}{z \, (1-z)} \, e^{- i \un \Delta \cdot \left[ z \, \un x_1 + (1-z) \, \un x_2\right]} \,  \\ 
& \times \, \sum_{\sigma_1, \sigma_2} \, \left\{  \Psi_{\lambda, \sigma_1, \sigma_2}^{\gamma^* \to q {\bar q}} ({\un x}_{02}, z) \, \left[ \Psi_{\lambda_V, \sigma_1, \sigma_2}^{V} ({\un x}_{12}, z) \right]^* \, \frac{1}{z} \, \llangle \tord \tr \left[ V^{\textrm{G} [2]}_{\un 1, \un 0} \, V_{\un 2}^\dagger \right] \rrangle (z s)   \right. \notag \\
& \hspace{1.5cm} \left. - \Psi_{\lambda, \sigma_1, \sigma_2}^{\gamma^* \to q {\bar q}} ({\un x}_{10}, z) \, \left[ \Psi_{\lambda_V, \sigma_1, \sigma_2}^{V} ({\un x}_{12}, z) \right]^* \, \frac{1}{1-z} \, \llangle \tord \tr \left[ V_{\un 1} \, V^{\textrm{G} [2] \, \dagger}_{\un 2, \un 0}  \right] \rrangle ((1-z) s)  \right\} . 
\label{A_subeik_2G}
\end{align}
Analogous to \eqs{type1_overlaps}, we define the following overlaps of the photon and meson wavefunctions relevant for the type-2 terms
\begin{subequations}\label{type2_overlaps}
\begin{align}
\notag
 \textrm{\bf TT \& T, -T:}& \hspace{0.5cm} \sum_{\sigma_1, \sigma_2} \, \Psi_{\lambda = T, \sigma_1, \sigma_2}^{\gamma^* \to q {\bar q}} ({\un x}, z) \, \left[ \Psi_{\lambda_V = T, \sigma_1, \sigma_2}^{V} ({\un x}_{12}, z) \right]^* 
\\ \notag & \hspace{1.5cm}
= \frac{e Z_f}{\pi}  \, \Bigg\{ - \left[ (1-2z)^2 + \lambda \, \lambda_V \right] \, a_f \, K_1 (x_{\perp} \, a_f) \, \frac{\pd \phi_{T} (x_{12},z)}{\pd x_{12}} \label{TTT-T} \\
& \hspace*{3cm} \times \left[ \delta_{\lambda \lambda_V} \, \frac{{\un x} \cdot {\un x}_{12} + i \, \lambda \, {\un x} \times {\un x}_{12}}{2} - \delta_{\lambda, - \lambda_V} \, \frac{x_\perp \, x_{12}}{2} \, e^{- i \lambda (\phi_x + \phi_{12}) } \right] \frac{1}{x_\perp x_{12}}
\notag
\\  & \hspace*{3cm}
+ \delta_{\lambda \lambda_V} \, m_f^2 \, K_0 (x_{\perp} \, a_f) \, \phi_{T} (x_{12},z) \Bigg\} \notag \\
& \hspace*{1.5cm}  \equiv \, \delta_{\lambda \lambda_V} \, \Phi^{\prime [2]}_{TT} ({\un x} , {\un x}_{12} , z) + i \, \lambda \, \delta_{\lambda \lambda_V} \,  \Phi^{[2]}_{TT} ({\un x} , {\un x}_{12} , z)  + \delta_{\lambda, - \lambda_V} \,  e^{- i \lambda (\phi_x + \phi_{12}) }  \, \Phi^{[2]}_{T, -T} ({x}_\perp , {x}_{12} , z) ,  \\
\notag 
\textrm{\bf TL:}& \hspace{0.5cm} \sum_{\sigma_1, \sigma_2} \, \Psi_{\lambda = T, \sigma_1, \sigma_2}^{\gamma^* \to q {\bar q}} ({\un x}, z) \, \left[ \Psi_{\lambda_V = 0, \sigma_1, \sigma_2}^{V} ({\un x}_{12}, z) \right]^* 
\\ \notag & \hspace{1.5cm}
= - \frac{e Z_f}{\pi} \, z (1-z) (1-2z) \, i \, a_f  \, \frac{{\un \epsilon}_{\lambda} \cdot {\un x}}{x_{\perp}} \, K_1 (x_{\perp} \, a_f)  \left[ M_V + \delta \, \frac{m_f^2 - \nabla_{12}^2}{ M_V \, z \, (1-z)} \right] \, \phi_{L}(x_{12},z) 
\\ & \hspace{1.5cm}
\equiv i \, \frac{{\un \epsilon}_{\lambda} \cdot {\un x}}{x_{\perp}} \,  \Phi^{[2]}_{TL} ({x}_\perp , {x}_{12} , z), \\
\notag
\textrm{\bf LT:}&\hspace{0.5cm} \sum_{\sigma_1, \sigma_2} \, \Psi_{\lambda = 0, \sigma_1, \sigma_2}^{\gamma^* \to q {\bar q}} ({\un x}, z) \, \left[ \Psi_{\lambda_V = T, \sigma_1, \sigma_2}^{V} ({\un x}_{12}, z) \right]^* 
\\ \notag & \hspace{1.5cm}
= - \frac{e Z_f}{\pi} \, z (1-z) (1-2z) \, i \, 2 Q \, K_0 (x_{\perp} \, a_f) \, \frac{{\un \epsilon}^*_{\lambda_V} \cdot {\un x}_{12}}{x_{12}} \, \frac{\pd \phi_{T} (x_{12},z)}{\pd x_{12}}  \notag 
\\  & \hspace{1.5cm}
\equiv i \, \frac{{\un \epsilon}^*_{\lambda_V} \cdot {\un x}_{12}}{x_{12}} \, \Phi^{[2]}_{LT} ({x}_\perp , {x}_{12} , z),  \\
\notag
\textrm{\bf LL:}&\hspace{0.5cm}  \sum_{\sigma_1, \sigma_2} \, \Psi_{\lambda = 0, \sigma_1, \sigma_2}^{\gamma^* \to q {\bar q}} ({\un x}, z) \, \left[ \Psi_{\lambda_V = 0, \sigma_1, \sigma_2}^{V} ({\un x}_{12}, z) \right]^* 
\\ \notag & \hspace{1.5cm}
= \frac{e Z_f}{\pi} \, z^2 (1-z)^2  \, 2 Q \, K_0 (x_{\perp} \, a_f) \, \left[ M_V + \delta \, \frac{m_f^2 - \nabla_{12}^2}{ M_V \, z \, (1-z)} \right] \, \phi_{L}(x_{12},z) 
\\  & \hspace{1.5cm}
\equiv \Phi^{[2]}_{LL} ({x}_\perp , {x}_{12} , z) .
\end{align}
\end{subequations}
Note, the structure in \eq{TTT-T} stems from 
\begin{align}
({\un \epsilon}_\lambda \cdot {\un v}) \ ({\un \epsilon}^*_{\lambda'} \cdot {\un w}) =  \delta_{\lambda, \lambda'} \, \left[ \frac{1}{2} \, {\un v} \cdot {\un w} + \frac{i}{2} \, \lambda \, {\un v} \times {\un w} \right] - \delta_{\lambda, - \lambda'} \, \frac{v_\perp \, w_\perp}{2} \, e^{- i \lambda (\phi_v + \phi_w) }
\end{align}
for arbitrary two-dimensional vectors $\un v$ and $\un w$. The cross product is defined by ${\un v} \times {\un w} = v^1 w^2 - v^2 w^1 = \epsilon^{ij} v^i w^j$ with $\epsilon^{ij}$ the two-dimensional Levi-Civita symbol and $i,j = 1,2$. $\phi_v$ and $\phi_w$ are the azimuthal angles of the two vectors. As mentioned above, explicit expressions for the wavefunction overlaps we will need below are given in Appendix~\ref{app:overlaps}.

%%%%%%%%%%%%%%%%%%%%%%%%%%%%%%%%%%%

\subsection{Type-3 operators}
\label{sec:type-3}

\begin{figure}[ht!]
    \centering
    \includegraphics[scale=0.3]{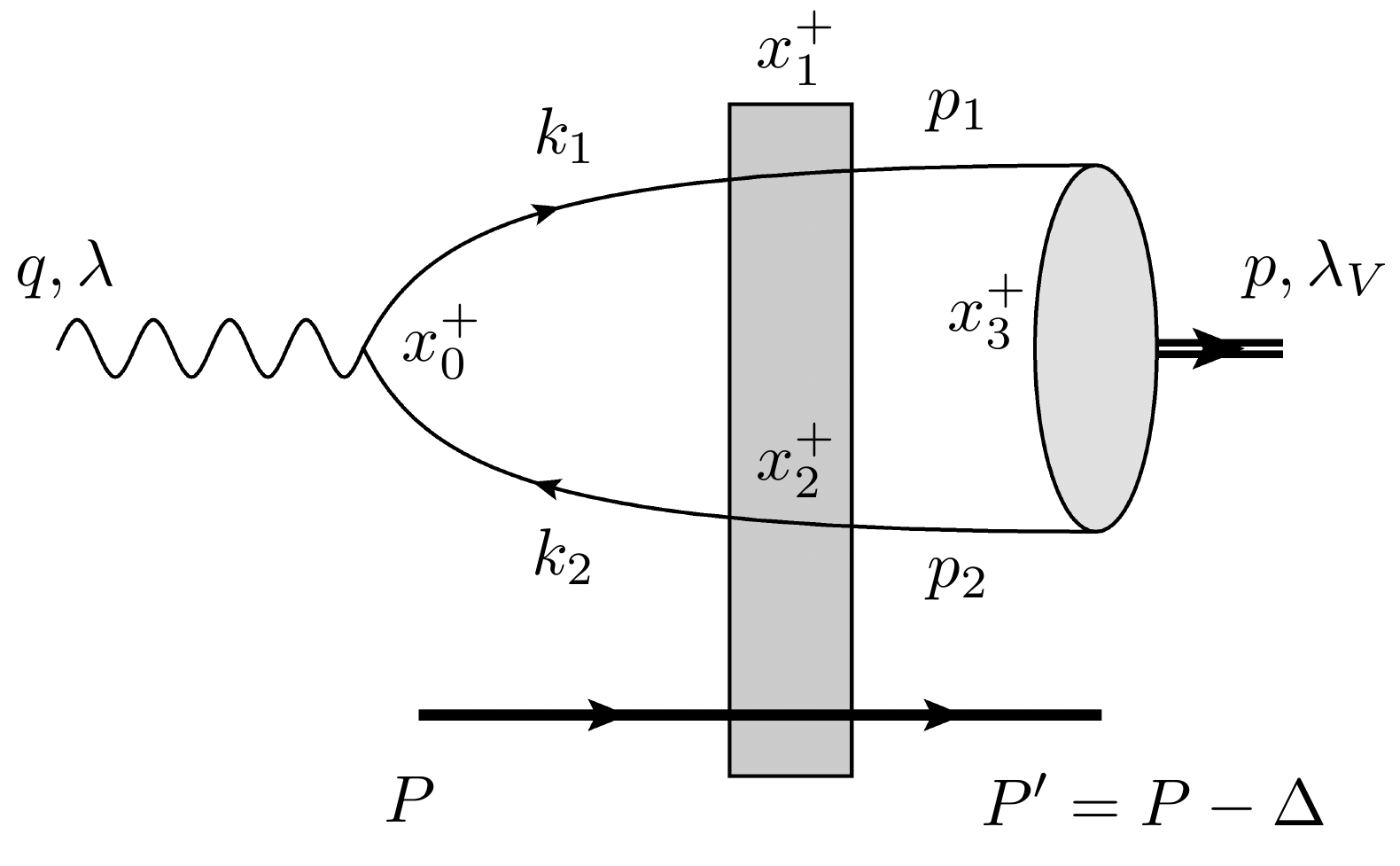}
    \caption{Diagram contributing to elastic vector meson production. The target shockwave is represented by the shaded rectangle, and the vector meson is denoted by the shaded oval.}
    \label{fig:vm_t3}
\end{figure}

For the type-3 operators, we first must temporarily undo the minus momentum conservation implied when simplifying \eq{imf}. To that end, consider the elastic vector meson amplitude depicted in \fig{fig:vm_t3}.
%%%%%%%%%%%%%%%%%%%%%%%%%%%%%%
%%%%%%%%%%%%%%%%%%%%%%%%%%%%%%

From \fig{fig:vm_t3}, the reduced amplitude can be written as 
\begin{align}\label{AeikF+-0}
& A_{\lambda_V, \lambda} \, 2 q^- \, (2 \pi) \delta (p^- - q^- - \Delta^-) =  i \, \int d^2 x_1 \, d^2 x_2 \, e^{- i \un \Delta \cdot (z_p \, \un x_1 + \bar z_p \, \un x_2)} \, \int \limits_{0}^{\infty} \frac{dk_1^-}{4\pi k_1^-} \frac{dk_2^-}{4\pi k_2^-} \, 2 q^- \, \int \limits_{0}^{\infty} \frac{dp_1^-}{4\pi p_1^-} \frac{dp_2^-}{4\pi p_2^-} \, 2 p^- \notag \\
& \times  \,
    \int \limits^{\infty}_{-\infty} dx_1^+ \, \int \limits^{\infty}_{-\infty} dx_2^+ \,  \int \limits^{\infty}_{-\infty} dx_0^+ \,  \int \limits^{\infty}_{-\infty} dx_3^+ 
     \, e^{i x_1^+ \left( p_1^- - k_1^- \right) + i x_2^+ \left( p_2^- - k_2^- \right)}  e^{i x_0^+ \left( k_1^- + k_2^-  - q^- \right) + i x_3^+ \left( p^- - p_1^-  - p_2^- \right) }  \, 2 \sqrt{k_1^- \, p_1^-} \, 2 \sqrt{k_2^- \, p_2^-} \notag \\  
     & \times \, \sum_{\sigma_1, \sigma_2} \Psi_{\lambda, \sigma_1, \sigma_2}^{\gamma^* \to q {\bar q}} \left( {\un x}_{12}, \frac{k_1^-}{q^-}, \frac{k_2^-}{q^-} \right) \, \left\langle \tord \tr \left[ 1 - V_{{\un 1}} (x_1^+)  \, V_{{\un 2}}^\dagger (x_2^+) \right] \right\rangle \, \left[ \Psi_{\lambda_V, \sigma_1, \sigma_2}^{V} \left({\un x}_{12}, \frac{p_1^-}{p^-}, \frac{p_2^-}{p^-} \right) \right]^*. 
\end{align}
where we have defined $z_p = p_1^-/p^-$, $ \bar z_p= p_2^-/p^-$ and explicitly show both the quark and anti-quark longitudinal momentum fractions in the arguments of the light-cone wave functions. We change the integration variables to 
\begin{align}
    y_1^+ \equiv x_1^+ - x_0^+, \ \ \ y_2^+ \equiv x_2^+ - x_0^+ 
\end{align}
and integrate out $x_0^+$ and $x_3^+$. Factoring out the overall delta function conserving minus momenta, we arrive at  
\begin{align}\label{AeikF+-2}
& A_{\lambda_V, \lambda} =  i \, \int d^2 x_1 \, d^2 x_2 \, e^{- i \un \Delta \cdot (z_p \, \un x_1 + \bar z_p \, \un x_2)} \, \int \limits_{0}^{\infty} \frac{dk_1^-}{2\pi} \frac{dk_2^-}{2\pi} \,  \int \limits_{0}^{p^-} \frac{dp_1^-}{4\pi p_1^-} \frac{p^-}{p^- -  p_1^-}  \notag \\
& \times  \,
    \int \limits^{\infty}_{-\infty} dy_1^+ \, \int \limits^{\infty}_{-\infty} dy_2^+  
     \, e^{i y_1^+ \left( p_1^- - k_1^- \right) + i y_2^+ \left( p^- -  p_1^- - k_2^- \right)}   \,  \sqrt{\frac{p_1^-}{k_1^-}} \, \sqrt{\frac{p^- -  p_1^-}{k_2^-}} \notag \\  
     & \times \, \sum_{\sigma_1, \sigma_2} \Psi_{\lambda, \sigma_1, \sigma_2}^{\gamma^* \to q {\bar q}} \left( {\un x}_{12}, \frac{k_1^-}{q^-}, \frac{k_2^-}{q^-} \right) \, \left\langle \tord \tr \left[ 1 - V_{{\un 1}} (y_1^+)  \, V_{{\un 2}}^\dagger (y_2^+) \right] \right\rangle \, \left[ \Psi_{\lambda_V, \sigma_1, \sigma_2}^{V} \left({\un x}_{12}, \frac{p_1^-}{p^-}, \frac{p^- -  p_1^-}{p^-} \right) \right]^*.
\end{align}
Expanding the Wilson lines in $y_1^+$ and $y_2^+$ around zero to the linear order and integrating these variables out yields for the resulting sub-eikonal part
\begin{align}\label{AeikF+-4}
& A^\textrm{[3] sub-eik}_{\lambda_V, \lambda} =  - \int d^2 x_1 \, d^2 x_2 \, e^{- i \un \Delta \cdot (z \, \un x_1 + (1- z) \, \un x_2)} \,  \int \limits_{0}^{p^-} \frac{dp_1^-}{4\pi p_1^-} \frac{1}{p^- -  p_1^-}  \, \sqrt{z (1-z)}  \,\sum_{\sigma_1, \sigma_2}
 \left[ \Psi_{\lambda_V, \sigma_1, \sigma_2}^{V} \left({\un x}_{12}, z, 1-z \right) \right]^*
\notag \\
& \times \, \Bigg\{  \left\langle \tord \tr \left[ \pd^- V_{{\un 1}} (0^+)  \, V_{{\un 2}}^\dagger (0^+) \, \frac{\pd}{\pd z} + V_{{\un 1}} (0^+)  \, \pd^-  V_{{\un 2}}^\dagger (0^+) \, \frac{\pd}{\pd \bar z} \right] \right\rangle \,
     \,  \frac{1}{\sqrt{z \, \bar z}} \, \Psi_{\lambda, \sigma_1, \sigma_2}^{\gamma^* \to q {\bar q}} \left( {\un x}_{12}, z, \bar z \right)  \Bigg\}_{\bar z = 1-z} 
\end{align}
where we have relabeled $z_p \to z$. In arriving at \eq{AeikF+-4}, we have assumed that the wave functions go to zero faster than $\sqrt{k_1^-}$ as $k_1^- = p_1^- \to 0$. We make a similar assumption as $k_2^- = p^- - p_1^- \to 0$. For models with $\phi_{L,T} (x_{12},z)$ behaving similar to \eq{bG}, e.g., for the ``boosted Gaussian" wave function \cite{Forshaw:2003ki, Kowalski:2006hc}, this is true. Furthermore, the energy dependence of the Wilson line correlators at $z=0$ and $z=1$ is likely a trivial $\sim 1/s$ behavior without any small-$x$ evolution enhancement and can be neglected compared to the terms we keep. Using \eq{VG3}, we write \eq{AeikF+-4} as
\begin{align}\label{AeikF+-6}
& A^\textrm{[3] sub-eik}_{\lambda_V, \lambda} =   \int d^2 x_1 \, d^2 x_2 \, e^{- i \un \Delta \cdot (z \, \un x_1 + (1- z) \, \un x_2)} \,  \int \limits_{0}^{1} \frac{dz}{2\pi z (1-z)}   \sum_{\sigma_1, \sigma_2}
 \left[ \Psi_{\lambda_V, \sigma_1, \sigma_2}^{V} \left({\un x}_{12}, z, 1-z \right) \right]^* \notag \\
& \times \, \Bigg\{  z \, \left\langle \tord \tr \left[ V_{{\un 1}}^{\mathrm{G} [3]}   \, V_{{\un 2}}^\dagger \right] \right\rangle \, \left( \frac{\pd}{\pd z} \Psi_{\lambda, \sigma_1, \sigma_2}^{\gamma^* \to q {\bar q}} \left( {\un x}_{12}, z, \bar z \right)  - \frac{1}{2 z} \, \Psi_{\lambda, \sigma_1, \sigma_2}^{\gamma^* \to q {\bar q}} \left( {\un x}_{12}, z, \bar z \right) \right) \notag \\
&  + \bar z \, \left\langle \tord \tr \left[ V_{{\un 1}} \, V_{{\un 2}}^{\mathrm{G} [3] \, \dagger} \right] \right\rangle \, \left( \frac{\pd}{\pd \bar z} \Psi_{\lambda, \sigma_1, \sigma_2}^{\gamma^* \to q {\bar q}} \left( {\un x}_{12}, z, \bar z \right)  - \frac{1}{2 \bar z} \, \Psi_{\lambda, \sigma_1, \sigma_2}^{\gamma^* \to q {\bar q}} \left( {\un x}_{12}, z, \bar z \right)  \right)  \Bigg\}_{\bar z = 1-z}.
\end{align}
In our sub-eikonal calculation, the following combination is sub-sub-eikonal and can be discarded due to the smallness of the longitudinal momentum transfer $\Delta^- = {\cal O} \left( 1/P^+ \right)$ by applying the argument from \cite{Becker:2026xcp}:
\begin{align}\label{zero_tr2}
    \llangle \tord \tr \left[  V_{\un 1}^{\mathrm{G} [3]}  \, V^\dagger_{\un 2} \right] \rrangle + \llangle \tord \tr \left[  V_{\un 1} \, V^{\mathrm{G} [3] \, \dagger}_{\un 2} \right] \rrangle = {\cal O} \left( \frac{1}{P^+} \right) \approx 0. 
\end{align}
We use this to simplify \eq{AeikF+-6} to 
\begin{align}\label{AeikF+-7}
& A^\textrm{[3] sub-eik}_{\lambda_V, \lambda} =  \frac{1}{4 \pi \, s} \,  \int d^2 x_1 \, d^2 x_2 \, e^{- i \un \Delta \cdot (z \, \un x_1 + (1- z) \, \un x_2)} \,  \int \limits_{0}^{1} \frac{dz}{z (1-z)} \,  \sum_{\sigma_1, \sigma_2}  \left[ \Psi_{\lambda_V, \sigma_1, \sigma_2}^{V} \left({\un x}_{12}, z \right) \right]^*   \\
& \times \, \Bigg\{  \llangle \tord \tr \left[ V_{{\un 1}}^{\mathrm{G} [3]}   \, V_{{\un 2}}^\dagger \right] \rrangle (z s) - \llangle \tord \tr \left[ V_{{\un 1}} \, V_{{\un 2}}^{\mathrm{G} [3] \, \dagger} \right] \rrangle ((1-z) s) \Bigg\} \, \left( \frac{\pd}{\pd z} \Psi_{\lambda, \sigma_1, \sigma_2}^{\gamma^* \to q {\bar q}} \left( {\un x}_{12}, z \right)  + \frac{z - \thalf}{z (1-z)} \, \Psi_{\lambda, \sigma_1, \sigma_2}^{\gamma^* \to q {\bar q}} \left( {\un x}_{12}, z \right) \right). \notag
\end{align}
We have returned to the conventional notation for the light-cone wave functions as functions of only one longitudinal momentum fraction $z$. Here $\partial/\partial z$ denotes the derivative of $\Psi^{\gamma^* \to q {\bar q}}_{\lambda, \sigma_1, \sigma_2} ( {\un x}_{12}, z ) \equiv \Psi^{\gamma^* \to q {\bar q}}_{\lambda, \sigma_1, \sigma_2} ( {\un x}_{12}, z, 1-z )$ with respect to $z$. As with the type-1 and type-2 terms in the previous Sections, we will use an abbreviated notation for the wavefunction overlaps appearing in \eq{AeikF+-7}. While the terms without the derivatives in $z$ can be described via the type-2 overlaps in \eqref{type2_overlaps}, the terms with the $z$-derivatives additionally require the following definitions
\begin{subequations}\label{type3_overlaps}
\begin{align}
\notag
 \textrm{\bf TT \& T, -T:}& \hspace{0.5cm} \sum_{\sigma_1, \sigma_2} \, \frac{\partial}{\partial z} \Big[ \Psi_{\lambda = T, \sigma_1, \sigma_2}^{\gamma^* \to q {\bar q}} ({\un x}_{12}, z) \Big] \, \left[ \Psi_{\lambda_V = T, \sigma_1, \sigma_2}^{V} ({\un x}_{12}, z) \right]^* 
\\ \notag & \hspace{-1cm}
= \frac{e Z_f}{\pi} \, (1-2z) \, \Bigg\{ \delta_{\lambda \lambda_V} \, \Bigg[ \frac{\pd \phi_{T} (x_{12},z)}{\pd x_{12}} \bigg( \frac{1 + (1-2z)^2}{4} \, x_{12} \, Q^2 \, K_0 (x_{12} \, a_f) 
- \frac{(1-2z)^2}{2 z (1-z)} \, a_f \, K_1 (x_{12} \, a_f) \bigg) 
\\ \notag & \hspace*{4cm}
+ m_f^2 \left( \frac{K_0 (x_{12} \, a_f)}{2 z (1-z)} - \frac{x_{12} \, Q^2}{2 \, a_f} \, K_1 (x_{12} \, a_f) \right) \phi_{T} (x_{12},z) \Bigg]
\\ \notag & \hspace*{3cm}
+ \delta_{\lambda, - \lambda_V} \, e^{- 2 i \lambda \phi_{12}} \, \Big[ z (1-z) \, x_{12} \, Q^2 \, K_0 (x_{12} \, a_f) - 2 \, a_f \, K_1 (x_{12} \, a_f) \Big] \frac{\pd \phi_{T} (x_{12},z)}{\pd x_{12}} \Bigg\} 
\\ & \hspace*{1.5cm}  \equiv \, \delta_{\lambda \lambda_V} \, \Phi^{\prime \, [3]}_{TT} ( x_{12} , z) + \delta_{\lambda, - \lambda_V} \,  e^{- 2 i \lambda \phi_{12} }  \, \Phi^{[3]}_{T, -T} ( x_{12} , z) ,  \\
\notag 
\textrm{\bf TL:}& \hspace{0.5cm} \sum_{\sigma_1, \sigma_2} \, \frac{\partial}{\partial z}  \Big[ \Psi_{\lambda = T, \sigma_1, \sigma_2}^{\gamma^* \to q {\bar q}} ({\un x}_{12}, z) \Big] \, \left[ \Psi_{\lambda_V = 0, \sigma_1, \sigma_2}^{V} ({\un x}_{12}, z) \right]^* 
\\ \notag & \hspace{1.5cm}
= - \frac{e Z_f}{2 \pi} \, i \, \frac{{\un \epsilon}_{\lambda} \cdot {\un x}_{12}}{x_{12}} \, \Big[ \big( 1 - 8 z (1-z) \big) \, a_f \, K_1 (x_{12} \, a_f) - z (1-z) (1-2z)^2 \, x_{12} \, Q^2 \, K_0 (x_{12} \, a_f) \Big] 
\\ \notag & \hspace*{3cm}
\times \left[ M_V + \delta \, \frac{m_f^2 - \nabla_{12}^2}{ M_V \, z \, (1-z)} \right] \, \phi_{L}(x_{12},z) 
\\ & \hspace{1.5cm}
\equiv i \, \frac{{\un \epsilon}_{\lambda} \cdot {\un x}_{12}}{x_{12}} \,  \Phi^{[3]}_{TL} ( x_{12} , z), \\
\notag
\textrm{\bf LT:}&\hspace{0.5cm} \sum_{\sigma_1, \sigma_2} \, \frac{\partial}{\partial z}  \Big[ \Psi_{\lambda = 0, \sigma_1, \sigma_2}^{\gamma^* \to q {\bar q}} ({\un x}_{12}, z) \Big] \, \left[ \Psi_{\lambda_V = T, \sigma_1, \sigma_2}^{V} ({\un x}_{12}, z) \right]^* 
\\ \notag & \hspace{1.5cm}
= - \frac{e Z_f}{2 \pi} \, i \, 2 Q \, (1-2z)^2 \, \Big[ 3 \, K_0 (x_{12} \, a_f) - \frac{z (1-z) \, x_{12} \, Q^2}{a_f} \, K_1 (x_{12} \, a_f) \Big] 
\\ \notag & \hspace*{3cm}
\times \frac{{\un \epsilon}^*_{\lambda_V} \cdot {\un x}_{12}}{x_{12}} \, \frac{\pd \phi_{T} (x_{12},z)}{\pd x_{12}}
\\  & \hspace{1.5cm}
\equiv - i \, \frac{{\un \epsilon}^*_{\lambda_V} \cdot {\un x}_{12}}{x_{12}} \, \Phi^{[3]}_{LT} ( x_{12} , z),  \\
\notag
\textrm{\bf LL:}&\hspace{0.5cm}  \sum_{\sigma_1, \sigma_2} \, \frac{\partial}{\partial z}  \Big[ \Psi_{\lambda = 0, \sigma_1, \sigma_2}^{\gamma^* \to q {\bar q}} ({\un x}_{12}, z) \Big] \, \left[ \Psi_{\lambda_V = 0, \sigma_1, \sigma_2}^{V} ({\un x}_{12}, z) \right]^* 
\\ \notag & \hspace{1.5cm}
= \frac{e Z_f}{2 \pi} \, 2 Q \, z (1-z) (1-2z) \, \Big[ 3 \, K_0 (x_{12} \, a_f) - \frac{z (1-z) \, x_{12} \, Q^2}{a_f} \, K_1 (x_{12} \, a_f) \Big] 
\\ \notag & \hspace*{3cm}
\times \left[ M_V + \delta \, \frac{m_f^2 - \nabla_{12}^2}{ M_V \, z \, (1-z)} \right] \, \phi_{L}(x_{12},z) 
\\  & \hspace{1.5cm}
\equiv \Phi^{[3]}_{LL} ( x_{12} , z) .
\end{align}
\end{subequations}
~\\

With the eikonal amplitude in \eq{A_eik3}, the sub-eikonal amplitudes in \eqs{A_subeik_11}, \eqref{A_subeik_2q}, \eqref{A_subeik_2G}, and \eqref{AeikF+-7}, and the associated wave function overlaps in \eqs{type1_overlaps}, \eqref{type2_overlaps}, and \eqref{type3_overlaps} in hand, we now proceed to compute the LT structure of the elastic vector meson DSA for transversely and longitudinally polarized vector mesons. Note that the full sub-eikonal amplitude for the elastic vector meson production is a sum of all the contributions we have found: 
\begin{align}
    A^\textrm{sub-eik}_{\lambda_V, \lambda} = A^\textrm{[1] sub-eik}_{\lambda_V, \lambda} + A^\textrm{\textrm{q}[2] sub-eik}_{\lambda_V, \lambda} + A^\textrm{\textrm{G}[2] sub-eik}_{\lambda_V, \lambda} + A^\textrm{[3] sub-eik}_{\lambda_V, \lambda} .
\end{align}

%%%%%%%%%%%%%%%%%%%%%%%%%%%%%%%%%%%

\section{Double spin asymmetry: transversely polarized vector mesons}
\label{sec:dsa_t}

We now evaluate the LT structure of \eq{dsa_t}, that is, the second term in its curly brackets, for a transversely polarized vector meson. As explained below \eqs{VMdsas}, this is the only structure in \eq{dsa_t} that is sensitive to OAM. Using \eq{XS_VM1}, it can be written as
\begin{align} \label{dsa_lt}
& \half \sum_{S_L = \pm 1} \!\!\!\! S_L \!\!\!\sum_{\lambda, \lambda_V = \pm 1} \left[ e^{i \lambda \phi} \frac{d\sigma_{0 \lambda; \lambda_V}^{\gamma^* p \to V p'}}{d^2 \Delta} + \cc \right] = \frac{1}{4 \pi^2} \sum_{\lambda, \lambda_V = \pm 1} \left\{ e^{i \lambda \phi} \left[ A^{\textrm{eik}}_{\lambda_V, \lambda} \left( A^{\textrm{sub-eik}}_{\lambda_V, 0} \right)^* + A^{\textrm{sub-eik}}_{\lambda_V, \lambda} \left( A^{\textrm{eik}}_{\lambda_V, 0} \right)^* \right] + \cc \right\} ,
\end{align}
where, as in \eq{dsa_t}, $\phi$ is the azimuthal angle of the incoming electron.\footnote{In the dipole frame, where the virtual photon has no transverse momentum, the outgoing and incoming electrons have the same azimuthal angle.} We also sum over the transverse polarizations of the vector meson. 

The sums over the photon and meson helicities will be carried out below with the help of
\begin{align} \label{sum_lambda_V}
\sum_{\lambda_V = \pm 1} e^{i \lambda_V \phi} \, \frac{{\un \epsilon}_{\lambda_V} \cdot {\un x}}{x_\perp} = i \, \sqrt{2} \, \sin (\phi - \phi_{x}), \qquad \sum_{\lambda_V = \pm 1} \lambda_V \, e^{i \lambda_V \phi} \, \frac{{\un \epsilon}_{\lambda_V} \cdot {\un x}}{x_\perp} = \sqrt{2} \, \cos (\phi - \phi_{x}) ,
\end{align}
with $\phi_x$ the azimuthal angle of $\un x$. 

%%%%%%%%%%%%%%%%%%%%%%%%%%%%%%%%%%%%

\subsection{Type-1 contribution}

We begin with the type-1 operators. Inserting \eqs{A_eik3} and \eqref{A_subeik_11} into \eq{dsa_lt}, employing the wavefunction overlaps in \eqs{type1_overlaps} along with the type-2 overlaps of \eqs{type2_overlaps} (for the eikonal amplitude), and summing over $\lambda$ and $\lambda_V$ with the help of \eq{sum_lambda_V}, we arrive at
\begin{align} \label{lt1_raw}
& \half \sum_{S_L = \pm 1} S_L \sum_{\lambda, \lambda_V = \pm 1} \left[ e^{i \lambda \phi} \frac{d\sigma_{0 \lambda; \lambda_V}^{\gamma^* p \to V p' \, [1]}}{d^2 \Delta} + \cc \right] = \frac{i \, N_c \sqrt{2}}{4 \pi^2 \, s} \, \int \frac{d^2 x_1 \, d^2 x_2 \, d^2 x_{1'} \, d^2 x_{2'}}{(4 \pi)^2} \int\limits_0^1 \frac{dz}{z \, (1-z)} \int\limits_0^1 \frac{dz'}{z' \, (1-z')} \notag \\
& \times e^{- i \un \Delta \cdot \left[ z \un x_{1} + (1-z) \un x_2 - z' \un x_{1'} - (1-z')\, \un x_{2'} \right]}  \, \Bigg\{ \frac{1}{z} \, \cos (\phi - \phi_{1'2'}) \left[ \Phi^{\prime \, [1]}_{TT} (x_{12}, z) - \Phi^{[1]}_{TT} (x_{12}, z) \right] \Phi^{[2]}_{LT} (x_{1'2'}, x_{1'2'}, z') \, N_{1'2'}(s) \notag \\
& \hspace{2.2cm} \times \llangle \tord \tr \left[ V^{\textrm{pol} [1]}_{\un 1} V_{\un 2}^\dagger \right] + \tord \tr \left[ V_{\un 2} V^{\textrm{pol} [1] \, \dagger}_{\un 1} \right] \rrangle (zs) \notag \\
& \hspace{0.6cm} - \frac{1}{z'} \Big[ \cos (\phi - \phi_{1'2'}) \, \Phi^{\prime [2]}_{TT} (\un x_{12}, \un x_{12}, z) + \cos (2 \phi_{12} - \phi - \phi_{1'2'}) \, \Phi^{[2]}_{T,-T} (x_{12}, x_{12}, z) \Big] \Phi^{[1]}_{LT} (x_{1'2'}, z') \notag \\
& \hspace{2.2cm} \times N_{12}(s) \, \llangle \tord \tr \left[ V^{\textrm{pol} [1]}_{\un 1'} V_{\un 2'}^\dagger \right] + \tord \tr \left[ V_{\un 2'} V^{\textrm{pol} [1] \, \dagger}_{\un 1'} \right] \rrangle^* (z's) \Bigg\} + \cc
\end{align}
In writing \eq{lt1_raw} we have already interchanged $\un x_1 \leftrightarrow \un x_2$ together with $z \leftrightarrow 1-z$ in the terms where the polarized Wilson line belongs to the antiquark, and likewise for the primed variables. Assuming that $\phi_{T,L}(x_\perp, z)$ are symmetric under $z \to 1-z$, one concludes that the overlaps $\Phi^{[1]}_{TT}$, $\Phi^{[2]}_{TL}$, $\Phi^{[2]}_{LT}$, $\Phi^{\prime \, [3]}_{TT}$, $\Phi^{[3]}_{T,-T}$, and $\Phi^{[3]}_{LL}$ are odd under $z \to 1-z$, while all the remaining overlaps are even. This is what allows the two terms in \eq{A_subeik_11} to be combined. In \eq{lt1_raw}, $\phi_{12}$ and $\phi_{1'2'}$ denote the azimuthal angles of the dipole separations $\un x_{12}$ and $\un x_{1'2'}$, respectively. 

In adding the complex conjugate in \eq{lt1_raw}, we simultaneously flip $\un \Delta \to - \un \Delta$, $\un k \to - \un k$. This operation is allowed since, in the dipole frame, after the integrals over the transverse positions are carried out, the cross section only depends on the two remaining transverse vectors $\un \Delta$ and $\un{k}$: therefore, it can depend either on $\un \Delta \cdot \un k$ or on $\un \Delta \times \un k$. Either structure is invariant under $\un \Delta \to - \un \Delta$, $\un k \to - \un k$ and so is the cross section. This can also be understood as a consequence of the azimuthal rotational invariance of the cross section. Defining the polarized dipole amplitude of the first type \cite{Kovchegov:2015pbl, Kovchegov:2016weo, Kovchegov:2018zeq, Cougoulic:2022gbk, Kovchegov:2019rrz, Kovchegov:2023yzd},
\begin{align} \label{Q_def}
    Q_{12}(zs) = \frac{1}{2 \, N_c } \mathrm{Re} \, \llangle \tord \tr \left[ V^{\textrm{pol} [1]}_{\un 1} \, V_{\un 2}^\dagger \right] + \tord \tr \left[ V_{\un 2} V^{\textrm{pol} [1] \,\dagger}_{\un 1} \right] \rrangle (zs),
\end{align}
and explicitly adding the complex conjugate in \eq{lt1_raw} while noting that the $\un k \to - \un k$ transformation amounts to replacing $\phi \to \phi + \pi$ in \eq{lt1_raw}, we obtain
\begin{align} \label{lt1_final}
& \half \sum_{S_L = \pm 1} S_L \sum_{\lambda, \lambda_V = \pm 1} \left[ e^{i \lambda \phi} \frac{d\sigma_{0 \lambda; \lambda_V}^{\gamma^* p \to V p' \, [1]}}{d^2 \Delta} + \cc \right] = \frac{i \, N_c^2 \sqrt{2}}{\pi^2 \, s} \, \int \frac{d^2 x_1 \, d^2 x_2 \, d^2 x_{1'} \, d^2 x_{2'}}{(4 \pi)^2} \int\limits_0^1 \frac{dz}{z \, (1-z)} \int\limits_0^1 \frac{dz'}{z' \, (1-z')} \, \frac{1}{z} \notag \\
& \times e^{- i \un \Delta \cdot \left[ z \un x_{1} + (1-z) \un x_2 - z' \un x_{1'} - (1-z')\, \un x_{2'} \right]}  \, N_{1'2'}(s) \, Q_{12} (z s) \, 
\notag \\ 
& \times 
\Big\{ \Big[ \cos (\phi - \phi_{12}) \, \Phi^{\prime \,[2]}_{TT} (\un x_{1'2'}, \un x_{1'2'}, z') + \cos (2 \phi_{1'2'} - \phi - \phi_{12}) \, \Phi^{[2]}_{T,-T} (x_{1'2'}, x_{1'2'}, z') \Big] \Phi^{[1]}_{LT} (x_{12}, z)\notag \\
& \hspace{2.2cm} + \cos (\phi - \phi_{1'2'}) \left[ \Phi^{\prime \, [1]}_{TT} (x_{12}, z) - \Phi^{[1]}_{TT} (x_{12}, z) \right] \Phi^{[2]}_{LT} (x_{1'2'}, x_{1'2'}, z') \Big\} .
\end{align}
In arriving at \eq{lt1_final}, we have interchanged $\un x_{1} \leftrightarrow \un x_{1'}, \un x_{2} \leftrightarrow \un x_{2'},$ and $z \leftrightarrow z'$ in the terms multiplying $N_{12}(s)$ in \eq{lt1_raw}. When interchanging these variables, we must flip $\un \Delta \to - \un \Delta$, $\hat k \to - \hat k$ to account for the sign change in the exponent. 

%%%%%%%%%%%%%%%%%%%%%%%%%%%%%%%%%%

\subsection{Type-2 contributions}

We next consider the type-2 operators, beginning with $V^{\mathrm{q}[2]}_{\un x}$. Under a passive parity and time-reversal (PT) transformation, correlators of Wilson lines are simply complex conjugated \cite{Kovchegov:2023yzd, Cougoulic:2022gbk}, since the accompanying reflection $\un x_i \to - \un x_i$ can be absorbed into the definition of the transverse integration variables. Combining \eqs{A_subeik_2q} and \eqref{A_eik3} in \eq{dsa_lt}, one readily finds that the $V^{\mathrm{q}[2]}_{\un x}$ terms are PT-odd. The DSA is PT-even, so these terms drop out.\footnote{An analogous statement was made about the contribution of $V^{\mathrm{q}[2]}_{\un x}$ to the DSA in elastic dijet production considered in \cite{Kovchegov:2024wjs}. We refer the reader to \cite{Kovchegov:2024wjs} for more details about this argument.}

The contribution of $V^{\mathrm{G}[2]}_{\un x, \un y}$ follows from \eqs{A_subeik_2G} and \eqref{A_eik3}. Employing the wavefunction overlaps from \eqs{type2_overlaps} and summing over $\lambda$ and $\lambda_V$, we obtain
\begin{align} \label{lt2_raw}
& \half \sum_{S_L = \pm 1} S_L \sum_{\lambda, \lambda_V = \pm 1} \left[ e^{i \lambda \phi} \frac{d\sigma_{0 \lambda; \lambda_V}^{\gamma^* p \to V p' \, \mathrm{G}[2]}}{d^2 \Delta} + \cc \right] = - \frac{N_c \sqrt{2}}{4 \pi^2 \, s} \, \int \frac{d^2 x_1 \, d^2 x_2 \, d^2 x_0 \, d^2 x_{1'} \, d^2 x_{2'}}{(4 \pi)^2} \int\limits_0^1 \frac{dz}{z \, (1-z)} \int\limits_0^1 \frac{dz'}{z' \, (1-z')} \notag \\
& \times e^{- i \un \Delta \cdot \left[ z \un x_{1} + (1-z) \un x_2 - z' \un x_{1'} - (1-z')\, \un x_{2'} \right]} \, \Bigg\{ \Big[ \sin (\phi - \phi_{1'2'}) \, \Phi^{\prime \,[2]}_{TT} (\un x_{12}, \un x_{12}, z) + \sin (2 \phi_{12} - \phi - \phi_{1'2'}) \, \Phi^{[2]}_{T,-T} (x_{12}, x_{12}, z) \Big] \, N_{12}(s) \notag \\
& \hspace{0.6cm} \times \Bigg[ \frac{1}{z'} \, \Phi^{[2]}_{LT} (x_{02'}, x_{1'2'}, z') \, \llangle \tord \tr \left[ V^{\textrm{G} [2]}_{\un 1', \un 0} V_{\un 2'}^\dagger \right] \rrangle^* (z' s) \notag \\
& \hspace{1.4cm} - \frac{1}{1-z'} \, \Phi^{[2]}_{LT} (x_{1'0}, x_{1'2'}, z') \, \llangle \tord \tr \left[ V_{\un 1'} V^{\textrm{G} [2] \, \dagger}_{\un 2', \un 0} \right] \rrangle^* ((1-z')s) \Bigg] \notag \\
& + \Phi^{[2]}_{LT} (x_{1'2'}, x_{1'2'}, z') \, N_{1'2'}(s) \, \Bigg[ \frac{1}{z} \Big( \sin (\phi - \phi_{1'2'}) \, \Phi^{\prime \,[2]}_{TT} (\un x_{02}, \un x_{12}, z) + \cos (\phi - \phi_{1'2'}) \, \Phi^{[2]}_{TT} (\un x_{02}, \un x_{12}, z) \notag \\
& \hspace{1.0cm} + \sin (\phi_{02} + \phi_{12} - \phi - \phi_{1'2'}) \, \Phi^{[2]}_{T,-T} (x_{02}, x_{12}, z) \Big) \llangle \tord \tr \left[ V^{\textrm{G} [2]}_{\un 1, \un 0} V_{\un 2}^\dagger \right] \rrangle (z s) \notag \\
& \hspace{1.0cm} - \frac{1}{1-z} \Big( \sin (\phi - \phi_{1'2'}) \, \Phi^{\prime [2]}_{TT} (\un x_{10}, \un x_{12}, z) + \cos (\phi - \phi_{1'2'}) \, \Phi^{[2]}_{TT} (\un x_{10}, \un x_{12}, z) \notag \\
& \hspace{1.8cm} + \sin (\phi_{10} + \phi_{12} - \phi - \phi_{1'2'}) \, \Phi^{[2]}_{T,-T} (x_{10}, x_{12}, z) \Big) \llangle \tord \tr \left[ V_{\un 1} V^{\textrm{G} [2] \, \dagger}_{\un 2, \un 0} \right] \rrangle ((1-z)s) \Bigg] \Bigg\} + \cc ,
\end{align}
To proceed, we employ the identity \cite{Kovchegov:2024wjs, Kovchegov:2024aus}
 \begin{align}
  \int d^2 x_1 \, d^2 x_0 \, f ({\un x}_1, {\un x}_0) \, V^{\textrm{G} [2]}_{{\un 1}, {\un 0}} &= - i \, \int d^2 x_1 \Big[ \left( \partial_1^i - \partial^i_0 \right) f(\un x_1, \un x_0) 
 \Big]_{\un x_0 = \un x_1}\, V_{\un 1}^{i \, \mathrm{G}[2]} \notag\\ \notag & \hspace{0cm}
 + \frac{i P^+}{2s} \int d^2 x_1 
 \Big[ \left( \partial_1^2 + \partial^2_0 \right) f(\un x_1, \un x_0) 
 \Big]_{\un x_0 = \un x_1}\,
 \int \limits^\infty_{-\infty} dz^-\, V_{\un 1}
  \\  & \hspace{0cm}
   - \frac{i P^+}{s} \int d^2 x_1  \int \displaylimits^\infty_{-\infty} dz^-  f(\un x_1, \un x_1) \left[ V_{\un 1}[\infty, z^-] \cev{D}^i(z^-, \un x_1) D^i(z^-, \un x_1) V_{\un 1}[z^-, -\infty] \right], \label{VG2sub} 
 \end{align}
 where we have defined the following polarized Wilson line \cite{Cougoulic:2022gbk}
\begin{align} \label{d-d_pw}
    V_{\un z}^{i\mathrm{G}[2]} = \frac{P^+}{2s} \int \displaylimits^\infty_{-\infty} dz^- V_{\un z}[\infty, z^-] \left[ D^i(z^-, \un z) - \cev{D}^i(z^-, \un z) \right] V_{\un z}[z^-, -\infty].
\end{align}
 
The second and third terms of \eq{VG2sub} lead to PT-odd contributions and can be discarded. When the derivatives in the first term act on the overlaps, it is convenient to separate their tensor structures from their scalar content by defining
\begin{align} \label{Phi_decomp}
\Phi_{TT}^{\prime [2]} ({\un x}, {\un y}, z) = {\un x} \cdot {\un y} \; \Phi_{TT}^{\prime [2A]} ( x_\perp, y_\perp, z) + \Phi_{TT}^{\prime [2B]} ( x_\perp, y_\perp, z), \qquad \Phi_{TT}^{[2]} ({\un x}, {\un y}, z) = ({\un x} \times {\un y}) \; \overline{\Phi}_{TT}^{[2]} ( x_\perp, y_\perp, z),
\end{align}
and to denote the derivatives with respect to the first and second arguments by
\begin{align} \label{deriv_not}
    \Phi^{(10)}(x_\perp, x_\perp, z) = \left[\frac{\partial}{\partial x_\perp} \Phi( x_\perp, y_\perp, z)  \right]_{\un y = \un x}, \qquad
    \Phi^{(01)}(x_\perp, x_\perp, z) = \left[\frac{\partial}{\partial y_\perp} \Phi( x_\perp, y_\perp, z)  \right]_{\un y = \un x}.
\end{align}
Adding the complex conjugate, interchanging the primed and unprimed variables so that the polarized correlators are given in terms of unprimed variables as in the previous subsection, and defining the polarized dipole amplitude of the second type \cite{Kovchegov:2015pbl, Kovchegov:2016weo, Kovchegov:2018zeq, Cougoulic:2022gbk, Kovchegov:2019rrz, Kovchegov:2023yzd},
\begin{align} \label{Gi_def}
    G^i_{12}(zs) &= \frac{1}{2N_c} \mathrm{Re}\,  \llangle \tord \tr \left[ V_{\un 2} V_{\un 1}^{i\mathrm{G}[2] \dagger} \right] + \tord \tr \left[ V_{\un 1}^{i\mathrm{G}[2]}  V_{\un 2}^\dagger \right] \rrangle(zs),
\end{align}
we arrive at
\begin{align} \label{lt2_final}
& \half \sum_{S_L = \pm 1} S_L \sum_{\lambda, \lambda_V = \pm 1} \left[ e^{i \lambda \phi} \frac{d\sigma_{0 \lambda; \lambda_V}^{\gamma^* p \to V p' \, \mathrm{G}[2]}}{d^2 \Delta} + \cc \right] = - \frac{i \, N_c^2 \sqrt{2}}{\pi^2 \, s} \, \int \frac{d^2 x_1 \, d^2 x_2 \, d^2 x_{1'} \, d^2 x_{2'}}{(4 \pi)^2} \int\limits_0^1 \frac{dz}{z \, (1-z)} \int\limits_0^1 \frac{dz'}{z' \, (1-z')} \, \frac{1}{z} \notag \\
& \times e^{- i \un \Delta \cdot \left[ z \un x_{1} + (1-z) \un x_2 - z' \un x_{1'} - (1-z')\, \un x_{2'} \right]}  \, N_{1'2'}(s) \, G^i_{12} (zs) \notag \\
& \times \Bigg\{ \Big[ \sin (\phi - \phi_{12}) \, \Phi^{\prime [2]}_{TT} (\un x_{1'2'}, \un x_{1'2'}, z') + \sin (2 \phi_{1'2'} - \phi - \phi_{12}) \, \Phi^{[2]}_{T,-T} (x_{1'2'}, x_{1'2'}, z') \Big] \notag \\
& \hspace{2.2cm} \times \left[ \hat x^i_{12} \left( \Phi^{[2] (01)}_{LT} (x_{12}, x_{12}, z) - \Phi^{[2] (10)}_{LT} (x_{12}, x_{12}, z) \right) - i z \, \Delta^i \, \Phi^{[2]}_{LT} (x_{12}, x_{12}, z) \right] \notag \\
& \hspace{0.6cm} + \Big[ \cos (\phi - \phi_{12}) \, \Phi^{\prime [2]}_{TT} (\un x_{1'2'}, \un x_{1'2'}, z') + \cos (2 \phi_{1'2'} - \phi - \phi_{12}) \, \Phi^{[2]}_{T,-T} (x_{1'2'}, x_{1'2'}, z') \Big] \, \frac{\epsilon^{ij} x^j_{12}}{x_{12}^2} \, \Phi^{[2]}_{LT} (x_{12}, x_{12}, z) \notag \\
& \hspace{0.6cm} + \Phi^{[2]}_{LT} (x_{1'2'}, x_{1'2'}, z') \bigg[ \sin (\phi - \phi_{1'2'}) \, \hat x^i_{12} \Big( x_{12}^2 \, \Phi^{\prime [2A] (01)}_{TT} (x_{12}, x_{12}, z) - x_{12}^2 \, \Phi^{\prime [2A] (10)}_{TT} (x_{12}, x_{12}, z) \notag \\
& \hspace{5.6cm} + \Phi^{\prime [2B] (01)}_{TT} (x_{12}, x_{12}, z) - \Phi^{\prime [2B] (10)}_{TT} (x_{12}, x_{12}, z) \Big) \notag \\
& \hspace{0.8cm} - 2 \cos (\phi - \phi_{1'2'}) \, \epsilon^{ij} x^j_{12} \, \overline{\Phi}^{[2]}_{TT} (x_{12}, x_{12}, z) \notag \\
& \hspace{0.8cm} + \sin (2 \phi_{12} - \phi - \phi_{1'2'}) \, \hat x^i_{12} \left( \Phi^{[2] (01)}_{T,-T} (x_{12}, x_{12}, z) - \Phi^{[2] (10)}_{T,-T} (x_{12}, x_{12}, z) \right) \notag \\
& \hspace{0.8cm} - i z \, \Delta^i \Big( \sin (\phi - \phi_{1'2'}) \, \Phi^{\prime [2]}_{TT} (\un x_{12}, \un x_{12}, z) + \sin (2 \phi_{12} - \phi - \phi_{1'2'}) \, \Phi^{[2]}_{T,-T} (x_{12}, x_{12}, z) \Big) \bigg] \Bigg\} .
\end{align}
In deriving \eq{lt2_final} we have used the fact that $\Phi^{[2]}_{LT}$ is odd under $z' \to 1-z'$ and that the interchange $\un x_{1'} \leftrightarrow \un x_{2'}$ takes $\hat x^i_{1'2'} \to - \hat x^i_{1'2'}$ and $\phi_{1'2'} \to \phi_{1'2'} + \pi$.

%%%%%%%%%%%%%%%%%%%%%%%%%%%%%%%%%%%

\subsection{Type-3 contribution}

Turning to the type-3 operators, we use \eqs{AeikF+-7} and \eqref{A_eik3} in \eq{dsa_lt}. The wavefunction overlaps are now those of \eqs{type3_overlaps}, for the terms with the $z$-derivative, and those of \eqs{type2_overlaps}, for the remaining terms. Summing over $\lambda$ and $\lambda_V$ gives
\begin{align} \label{lt3_raw}
& \half \sum_{S_L=\pm1} S_L \sum_{\lambda, \lambda_V = \pm 1} \left[ e^{i \lambda \phi} \frac{d\sigma_{0 \lambda; \lambda_V}^{\gamma^* p \to V p' \, [3]}}{d^2 \Delta} + \cc \right] = \frac{i \, N_c \sqrt{2}}{4 \pi^2 \, s} \, \int \frac{d^2 x_1 \, d^2 x_2 \, d^2 x_{1'} \, d^2 x_{2'}}{(4 \pi)^2} \int\limits_0^1 \frac{dz}{z \, (1-z)} \int\limits_0^1 \frac{dz'}{z' \, (1-z')} \notag \\
& \times e^{- i \un \Delta \cdot \left[ z \un x_{1} + (1-z) \un x_2 - z' \un x_{1'} - (1-z')\, \un x_{2'} \right]}  \, \Bigg\{ - \Big[ \sin (\phi - \phi_{1'2'}) \, \Phi^{\prime [2]}_{TT} (\un x_{12}, \un x_{12}, z) + \sin (2 \phi_{12} - \phi - \phi_{1'2'}) \, \Phi^{[2]}_{T,-T} (x_{12}, x_{12}, z) \Big] \notag \\
& \hspace{1.4cm} \times \left[ \Phi^{[3]}_{LT} (x_{1'2'}, z') - \frac{z' - \thalf}{z' (1-z')} \, \Phi^{[2]}_{LT} (x_{1'2'}, x_{1'2'}, z') \right] \notag \\
& \hspace{2.2cm} \times N_{12}(s) \, \left[ \llangle \atord \tr \left[ V_{\un 2'} V^{\textrm{G} [3] \, \dagger}_{\un 1'} \right] \rrangle (z' s) - \llangle \atord \tr \left[ V^{\textrm{G} [3]}_{\un 2'} V^\dagger_{\un 1'} \right] \rrangle ((1-z') s) \right] \notag \\
& \hspace{0.6cm} - \Phi^{[2]}_{LT} (x_{1'2'}, x_{1'2'}, z') \, \bigg( \sin (\phi - \phi_{1'2'}) \left[ \Phi^{\prime \, [3]}_{TT} (x_{12}, z) + \frac{z - \thalf}{z (1-z)} \, \Phi^{\prime [2]}_{TT} (\un x_{12}, \un x_{12}, z) \right] \notag \\
& \hspace{3.0cm} + \sin (2 \phi_{12} - \phi - \phi_{1'2'}) \left[ \Phi^{[3]}_{T,-T} (x_{12}, z) + \frac{z - \thalf}{z (1-z)} \, \Phi^{[2]}_{T,-T} (x_{12}, x_{12}, z) \right] \bigg) \notag \\
& \hspace{2.2cm} \times N_{1'2'}(s) \, \left[ \llangle \tord \tr \left[ V^{\textrm{G} [3]}_{\un 1} V^\dagger_{\un 2} \right] \rrangle (zs) - \llangle \tord \tr \left[ V_{\un 1} V^{\textrm{G} [3] \, \dagger}_{\un 2} \right] \rrangle  ((1-z) s) \right] \Bigg\} + \cc .
\end{align}
 In the term containing $V^{\textrm{G} [3]}_{\un 2'}$ we interchange $\un x_{1'} \leftrightarrow \un x_{2'}$ together with $z' \leftrightarrow 1 - z'$, and in the term containing $V^{\textrm{G} [3] \, \dagger}_{\un 2}$ we interchange $\un x_{1} \leftrightarrow \un x_{2}$ with $z \leftrightarrow 1-z$. Adding the complex conjugate, interchanging the primed and unprimed variables so that the polarized correlators are given in terms of unprimed variables, and defining the polarized dipole amplitude of the third type \cite{Kovchegov:2025gcg, Becker:2026xcp},
\begin{align} \label{G3_def}
     G^{[3]}_{12}(zs) &= \frac{1}{2N_c} \mathrm{Re}\,  \llangle \tord \tr \left[ V_{\un 2} V_{\un 1}^{\mathrm{G}[3] \dagger} \right] + \tord \tr \left[ V_{\un 1}^{\mathrm{G}[3]}  V_{\un 2}^\dagger \right] \rrangle(zs),
\end{align}
we obtain
\begin{align} \label{lt3_final}
& \half \sum_{S_L = \pm 1} S_L \sum_{\lambda, \lambda_V = \pm 1} \left[ e^{i \lambda \phi} \frac{d\sigma_{0 \lambda; \lambda_V}^{\gamma^* p \to V p' \, [3]}}{d^2 \Delta} + \cc \right] = \frac{i \, N_c^2 \sqrt{2}}{\pi^2 \, s} \, \int \frac{d^2 x_1 \, d^2 x_2 \, d^2 x_{1'} \, d^2 x_{2'}}{(4 \pi)^2} \int\limits_0^1 \frac{dz}{z \, (1-z)} \int\limits_0^1 \frac{dz'}{z' \, (1-z')} \notag \\
& \times e^{- i \un \Delta \cdot \left[ z \un x_{1} + (1-z) \un x_2 - z' \un x_{1'} - (1-z')\, \un x_{2'} \right]}  \, N_{1'2'}(s) \, G^{[3]}_{12} (zs) \notag \\
& \times \Bigg\{ \Big[ \sin (\phi - \phi_{12}) \, \Phi^{\prime [2]}_{TT} (\un x_{1'2'}, \un x_{1'2'}, z') + \sin (2 \phi_{1'2'} - \phi - \phi_{12}) \, \Phi^{[2]}_{T,-T} (x_{1'2'}, x_{1'2'}, z') \Big] \notag \\
& \hspace{1.4cm} \times \left[ \Phi^{[3]}_{LT} (x_{12}, z) - \frac{z - \thalf}{z (1-z)} \, \Phi^{[2]}_{LT} (x_{12}, x_{12}, z) \right] \notag \\
& \hspace{0.6cm} - \Phi^{[2]}_{LT} (x_{1'2'}, x_{1'2'}, z') \bigg( \sin (\phi - \phi_{1'2'}) \left[ \Phi^{\prime \, [3]}_{TT} (x_{12}, z) + \frac{z - \thalf}{z (1-z)} \, \Phi^{\prime [2]}_{TT} (\un x_{12}, \un x_{12}, z) \right] \notag \\
& \hspace{3.0cm} + \sin (2 \phi_{12} - \phi - \phi_{1'2'}) \left[ \Phi^{[3]}_{T,-T} (x_{12}, z) + \frac{z - \thalf}{z (1-z)} \, \Phi^{[2]}_{T,-T} (x_{12}, x_{12}, z) \right] \bigg) \Bigg\} .
\end{align}
The complete LT structure for a transversely polarized vector meson is the sum of \eqs{lt1_final}, \eqref{lt2_final}, and \eqref{lt3_final}.

%%%%%%%%%%%%%%%%%%%%%%%%%%%%%%%%%%%

\subsection{Expansion in transverse momentum transfer $\Delta_\perp$}
\label{sec:dsat_expansion}

The moment amplitudes that determine the OAM distributions at small $x$ \cite{Kovchegov:2023yzd, Kovchegov:2024wjs, Becker:2026xcp} are first impact-parameter moments of the polarized dipole amplitudes. To access them, and, hence, the OAM distributions, we expand the LT structure to the linear order in $\un \Delta$. Defining the impact parameters 
\begin{align}
    \un b = z \, \un x_1 + (1-z) \, \un x_2, \qquad 
    \un b' = z' \, \un x_{1'} + (1-z') \, \un x_{2'},
\end{align}
and changing the integration variables to $d^2 x_1 \, d^2 x_2 \, d^2 x_{1'} \, d^2 x_{2'} = d^2 x_{12} \, d^2 x_{1'2'} \, d^2 b \, d^2 b'$, we expand the Fourier phase,
\begin{align}
e^{- i \un \Delta \cdot \left[ z \un x_{1} + (1-z) \un x_2 - z' \un x_{1'} - (1-z')\, \un x_{2'} \right]}  = e^{- i \un \Delta \cdot ({\un b} - {\un b}')} = 1 - i \, {\un \Delta} \cdot ({\un b} - {\un b}') + {\cal O} (\Delta_\perp^2).
\end{align}
To perform the integrals over impact-parameters, we need the impact-parameter integrated dipole amplitudes, defined as \cite{Kovchegov:2015pbl, Cougoulic:2022gbk, Becker:2026xcp}
\begin{subequations} \label{int_defs}
\begin{align}
 \int d^2 x_1 \, Q_{12} (zs) &= Q(x_{12}^2, zs), \\
 \int d^2 x_1 \, G^i_{12}(zs) &= x_{12}^i \, G_1(x_{12}^2, zs) + \epsilon^{ij} x_{12}^j \, G_2(x_{12}^2, zs),
 \\ 
 \int d^2 x_1\, G^{[3]}_{12}(zs) &= G^{[3]}(x_{12}^2, zs), 
\end{align}
\end{subequations}
along with their first impact-parameter moments \cite{Kovchegov:2023yzd, Kovchegov:2024wjs, Becker:2026xcp}
\begin{subequations} \label{mom_defs}
\begin{align}
 \int d^2 x_1 \,x_1^m\, Q_{12} (zs) &= x_{12}^m \,I_3(x_{12}^2, zs) + \epsilon^{mj} x_{12}^j \,J_3(x_{12}^2, zs), \\
 \int d^2 x_1 \,x_1^m\, G^i_{12}(zs) &= \epsilon^{mi} x_{12}^2 \, I_4(x_{12}^2, zs) + \epsilon^{mj} x_{12}^j x_{12}^i \, I_5(x_{12}^2, zs) \notag \\
 & \quad + \delta^{mi} x_{12}^2 \, J_4(x_{12}^2, zs) +  x_{12}^m x_{12}^i \, J_5(x_{12}^2, zs),
 \\ 
 \int d^2 x_1\, x_1^m \, G^{[3]}_{12}(zs) &= \epsilon^{mj} x_{12}^j \, I_E(x_{12}^2, zs) + x_{12}^m \, J_E(x_{12}^2, zs).
\end{align}
\end{subequations}
The impact-parameter integrals of the unpolarized dipole amplitude are
\begin{subequations} \label{n_ints}
\begin{align} 
 &\int d^2 x_1 \, N_{12}(s) = N(x_{12}^2, s),  \label{Nint} \\ 
 &\int d^2 x_1 \, \frac{x^m_1 + x^m_2}{2} \, N_{12}(s) = 0 .
\end{align}
\end{subequations}
Equations \eqref{int_defs}, \eqref{mom_defs}, and \eqref{n_ints} lead to 
\begin{subequations} \label{b_ints}
\begin{align}
& \int d^2 b \, b^i \, N_{12} (s) = \left( z - \thalf \right) \, x_{12}^i \, N (x_{12}^2, s), \\
& \int d^2 b \, b^i \, Q_{12}(zs) = {x}_{12}^i \, I_3 (x_{12}^2, zs) + \epsilon^{ij} x_{12}^j \, J_3 (x_{12}^2, zs) - (1-z) \, x_{12}^i \, Q (x_{12}^2, zs) , \\
& \int d^2 b \, b^i \, G^j_{12}(zs) = \epsilon^{ij} \, x_{12}^2 \, I_{4}(x_{12}^2, zs) + \epsilon^{ik} \, x_{12}^k \, x_{12}^j \, I_{5}(x_{12}^2 , zs) + \delta^{ij} x_{12}^2 \, J_{4}(x_{12}^2, zs) + x_{12}^i \, x_{12}^j \, J_{5}(x_{12}^2 , zs) \notag\\
& \hspace*{5cm} - (1-z) \, x_{12}^i \, \left[ x_{12}^j \, G_1(x_{12}^2, zs) + \epsilon^{jl} x_{12}^l \, G_2(x_{12}^2, zs) \right] ,
\\ 
& \int d^2b \, b^i \, G^{[3]}_{12}(zs) = \epsilon^{ij} x_{12}^j \, I_E(x_{12}^2, zs) + x_{12}^i \, J_E(x_{12}^2, zs) - (1-z) \, x_{12}^i \, G^{[3]}(x_{12}^2, zs).
\end{align}
\end{subequations}

We now expand \eqs{lt1_final}, \eqref{lt2_final}, and \eqref{lt3_final} to linear order in $\Delta_\perp$, integrate over the impact parameters with the help of \eqs{int_defs}, \eqref{Nint}, and \eqref{b_ints}, and then integrate over the dipole orientations $\phi_{12}$ and $\phi_{1'2'}$. At zeroth order in $\Delta_\perp$ everything vanishes after the angular integrals, as it must, since the LT structure is linear in $\hat k = \un k /k_\perp$. At the linear order in $\Delta_\perp$, two azimuthal harmonics survive, $\un \Delta \cdot \hat k \propto \cos (\phi - \phi_\Delta)$ and $\un \Delta \times \hat k \propto \sin (\phi - \phi_\Delta)$ (with $\phi_\Delta$ the azimuthal angle of $\un \Delta$), which can be separated experimentally by weighting the data with the appropriate trigonometric function of the azimuthal angle between the (outgoing) electron and the recoiling proton (in the dipole/Trento frame). The $\un \Delta \times \hat{k}$ structure descends from the more general structure $\hat{P} \cdot (\vec{\Delta} \times \vec{k})$, where $\vec{\Delta}$ and $\vec{k}$ are the three-vector parts of $\Delta^\mu = P^\mu - P'^\mu$ and $k^\mu$, respectively, while $\hat{P}$ is the unit vector along the 3-momentum of the incoming proton. The structure $\hat{P} \cdot (\vec{\Delta} \times \vec{k})$ is parity-odd and therefore inconsistent with the parity-even DSA. We thus expect the $\un \Delta \times \hat k$ harmonic to vanish identically once all the symmetry properties of the polarized dipole amplitudes are taken into account. Its coefficient involves the $J$-type moment amplitudes defined in \eqs{mom_defs} along with $G_1$ and $G^{[3]}$, which do not contribute to the OAM distributions we are interested in (a similar argument was made in \cite{Kovchegov:2024wjs}). Therefore, we do not display this harmonic here. 

Isolating the $\cos (\phi - \phi_\Delta)$ harmonic in the order-$\Delta_\perp$ terms gives us our main result for transversely polarized vector mesons,
\begin{tcolorbox}
\begin{align} \label{lt_exp_cos}
& \half \sum_{S_L = \pm 1} S_L \sum_{\lambda, \lambda_V = \pm 1} \left[ e^{i \lambda \phi} \frac{d\sigma_{0 \lambda; \lambda_V}^{\gamma^* p \to V p'}}{d^2 \Delta} + \cc \right]_{\cos (\phi - \phi_\Delta)} \notag \\
& = \un \Delta \cdot \hat{k} \; \frac{2 \, N_c^2 \sqrt{2}}{\pi^2 \, s} \int \frac{d x^2_{12} \, d x^2_{1'2'}}{64} \int\limits_0^1 \frac{dz}{z \, (1-z)} \int\limits_0^1 \frac{dz'}{z' \, (1-z')} \, \frac{1}{z} \, N (x^2_{1'2'} , s) \notag \\
& \times \Bigg\{ \Phi^{\prime [2]}_{TT} (\un x_{1'2'}, \un x_{1'2'}, z') \bigg[ x_{12}^2 \left( \Phi^{[2] (01)}_{LT} (x_{12}, x_{12}, z) - \Phi^{[2] (10)}_{LT} (x_{12}, x_{12}, z) \right) \Big( I_4 (x_{12}^2, zs) + I_5 (x_{12}^2, zs) \Big) \notag \\
& \hspace{2.6cm} + x_{12} \, \Phi^{[2]}_{LT} (x_{12}, x_{12}, z) \Big( I_4 (x_{12}^2, zs) + G_2 (x_{12}^2, zs) \Big) \notag \\
& \hspace{2.6cm} + x_{12} \, \Phi^{[1]}_{LT} (x_{12}, z) \Big( I_3 (x_{12}^2, zs) - (1-z) \, Q (x_{12}^2, zs) \Big) \notag \\
& \hspace{2.6cm} - z \, x_{12} \left( \Phi^{[3]}_{LT} (x_{12}, z) - \frac{z - \thalf}{z (1-z)} \, \Phi^{[2]}_{LT} (x_{12}, x_{12}, z) \right) I_E (x_{12}^2, zs) \bigg] \notag \\
& \hspace{0.6cm} + \left( z' - \thalf \right) x_{1'2'} \, \Phi^{[2]}_{LT} (x_{1'2'}, x_{1'2'}, z') \notag \\
& \hspace{2.2cm} \times \left[ \left( \Phi^{[1]}_{TT} (x_{12}, z) - \Phi^{\prime \, [1]}_{TT} (x_{12}, z) \right) Q (x_{12}^2, zs) - 2 \, x_{12}^2 \, \overline{\Phi}^{ [2]}_{TT} (x_{12}, x_{12}, z) \, G_2 (x_{12}^2, zs) \right] \Bigg\} + {\cal O} (\Delta_\perp^2),
\end{align}
\end{tcolorbox}
\noindent where the explicit expressions for the wavefunction overlaps are given in Appendix~\ref{app:overlaps}. Although we have written \eq{lt_exp_cos} for a single quark flavor $f$, it is understood that one needs to sum over the flavor content of the corresponding vector meson. For example, for the $\rho$ meson, one should form the combination $d\sigma^{V= \rho} = (1/\sqrt{2})( d\sigma_{f = u} - d\sigma_{f=d} )$, where $d\sigma_{f}$ is the cross section for a single flavor $f$. In phenomenological applications of \eq{lt_exp_cos}, the dipole amplitude $Q$ and the moment-amplitude $I_3$ should be furnished with a flavor index and would enter the flavor sum above accordingly.  

Equation \eqref{lt_exp_cos} contains $Q$ and $G_2$, which are constrained by the existing small-$x$ helicity phenomenology \cite{Adamiak:2021ppq, Adamiak:2023yhz, JAMCollaborationSmall-xAnalysisGroup:2025tfa}, and the Pomeron amplitude $N$, which is fixed by unpolarized data. Every ingredient of \eq{lt_exp_cos} that is not a moment amplitude is therefore known. We conclude that \eq{lt_exp_cos} may allow one to constrain, possibly from the future EIC data, the moment amplitudes $I_3$, $I_4$, $I_5$ (from the type-1 and type-2 operators) and $I_E$ (from the type-3 operator), which govern the OAM distributions at small $x$ \cite{Kovchegov:2023yzd, Kovchegov:2024wjs, Becker:2026xcp}. More specifically, the quark and gluon OAM distributions at small $x$ are given by \cite{Kovchegov:2023yzd, Kovchegov:2024wjs, Becker:2026xcp}
\begin{subequations}\label{OAMs_corr}
\begin{align}
        \label{final_quark_OAM_corr}
    & L_{q+\bar{q}}(x, Q^2) = \frac{2\,N_f}{\as \pi^2} \, \widetilde{I} \left(x_{10}^2 = \frac{1}{Q^2}, s = \frac{Q^2}{x} \right)  ,
    \\  \label{final_gluon_OAM}
    & L_G(x,Q^2) = - \frac{2 \, N_c}{\as \pi^2} 
    \Big[ 2\, I_4 + 3\, I_5\Big]\!\!\left(x_{10}^2 = \frac{1}{Q^2}, \, s = \frac{Q^2}{x} \right).
\end{align}
\end{subequations}
Here $\widetilde{I}$ is another moment-amplitude defined in \cite{Becker:2026xcp}. Its small-$x$ evolution in the double-logarithmic approximation (DLA) mixes it with $I_3$, $I_4$, $I_5$, $Q$, and $G_2$ \cite{Becker:2026xcp}. However, it is an independent degree of freedom which cannot be completely fixed by the other moment-amplitudes and dipole amplitudes, and, therefore by our suggested observable in \eq{lt_exp_cos}. Therefore, the quark OAM distribution $L_{q+\bar{q}}(x, Q^2)$ cannot be completely constrained by the elastic vector meson production process we consider here. On the positive side, \eq{lt_exp_cos} demonstrates that it is possible, at least in principle, to directly constrain $I_4$ and $I_5$, and, hence, via \eq{final_gluon_OAM}, the gluon OAM distribution $L_G(x,Q^2)$ by measuring the $\cos (\phi - \phi_{\Delta})$ harmonic of  the DSA in elastic transversely-polarized vector meson production at the EIC.

%%%%%%%%%%%%%%%%%%%%%%%%%%%%%%%%%%%%%%%%%%%%%%%%%%%%%%%%%%%%%%%%%%%%%%%%%%%%%%%

\section{Double spin asymmetry: longitudinally polarized vector mesons}
\label{sec:dsa_l}

We now repeat the analysis of Section~\ref{sec:dsa_t} for $\lambda_V = 0$, that is, for a longitudinally polarized vector meson. The relevant structure is the LT term of \eq{dsa_l},
\begin{align}
& \half \sum_{S_L = \pm 1} S_L \sum_{\lambda = \pm 1} \left[ e^{i \lambda \phi} \frac{d\sigma_{0 \lambda; \lambda_V = 0}^{\gamma^* p \to V p'}}{d^2 \Delta} + \cc \right]= \frac{1}{4 \pi^2} \sum_{\lambda = \pm 1} \left\{ e^{i \lambda \phi} \left[ A^{\textrm{eik}}_{0, \lambda} \left( A^{\textrm{sub-eik}}_{0, 0} \right)^* + A^{\textrm{sub-eik}}_{0, \lambda} \left( A^{\textrm{eik}}_{0, 0} \right)^* \right] + \cc \right\} . \label{dsal_lt}
\end{align}
Every step below is nearly identical to the transversely-polarized vector meson production case, and we only indicate where the two calculations differ.

%%%%%%%%%%%%%%%%%%%%%%%%%%%%%%%%%%%%%%%%%%%

\subsection{Type-1 contribution}
\label{sec:dsal_1}

Using \eqs{A_eik3} and \eqref{A_subeik_11} along with the wavefunction overlaps in \eqs{type1_overlaps} and \eqref{type2_overlaps}, and summing over $\lambda$, we find
\begin{align} \label{lt1_l_raw}
& \half \sum_{S_L = \pm 1} S_L \sum_{\lambda = \pm 1} \left[ e^{i \lambda \phi} \frac{d\sigma_{0 \lambda; \lambda_V = 0}^{\gamma^* p \to V p' \, [1]}}{d^2 \Delta} + \cc \right] = - \frac{i \, N_c \sqrt{2}}{4 \pi^2 \, s} \, \int \frac{d^2 x_1 \, d^2 x_2 \, d^2 x_{1'} \, d^2 x_{2'}}{(4 \pi)^2} \int\limits_0^1 \frac{dz}{z \, (1-z)} \int\limits_0^1 \frac{dz'}{z' \, (1-z')} \notag \\
& \times e^{- i \un \Delta \cdot \left[ z \un x_{1} + (1-z) \un x_2 - z' \un x_{1'} - (1-z')\, \un x_{2'} \right]}  \, \cos (\phi - \phi_{12}) \, \Phi^{[1]}_{TL} (x_{12}, z) \, \Phi^{[2]}_{LL} (x_{1'2'}, x_{1'2'}, z') \, N_{1'2'} (s) \notag \\
& \times \left[ - \frac{1}{z} \, \llangle \tord \tr \left[ V^{\textrm{pol} [1]}_{\un 1} V_{\un 2}^\dagger \right] \rrangle (z s) + \frac{1}{1-z} \, \llangle \tord \tr \left[ V_{\un 1} V^{\textrm{pol} [1] \, \dagger}_{\un 2} \right] \rrangle ((1-z) s) \right] + \cc .
\end{align}
The term with the type-1 operator in the complex conjugate amplitude $\left( A^{\textrm{sub-eik}}_{0, 0} \right)^*$ does not contribute, because it involves the LL type-1 overlap from \eq{LL_type-1}, which vanishes. Since $\Phi^{[1]}_{TL}$ is even under $\un x_1 \leftrightarrow \un x_2$, $z \leftrightarrow 1-z$ while $\cos (\phi - \phi_{12})$ is odd, this interchange combines the two traces into the correlator of \eq{Q_def}. Adding the complex conjugate, we obtain
\begin{align} \label{lt1_l}
& \half \sum_{S_L = \pm 1} S_L \sum_{\lambda = \pm 1} \left[ e^{i \lambda \phi} \frac{d\sigma_{0 \lambda; \lambda_V = 0}^{\gamma^* p \to V p' \, [1]}}{d^2 \Delta} + \cc \right] = \frac{i \, N_c^2 \sqrt{2}}{\pi^2 \, s} \, \int \frac{d^2 x_1 \, d^2 x_2 \, d^2 x_{1'} \, d^2 x_{2'}}{(4 \pi)^2} \int\limits_0^1 \frac{dz}{z \, (1-z)} \int\limits_0^1 \frac{dz'}{z' \, (1-z')} \, \frac{1}{z} \notag \\
& \times e^{- i \un \Delta \cdot \left[ z \un x_{1} + (1-z) \un x_2 - z' \un x_{1'} - (1-z')\, \un x_{2'} \right]}  \, N_{1'2'} (s) \, Q_{12} (zs) \, \cos (\phi - \phi_{12}) \, \Phi^{[1]}_{TL} (x_{12}, z) \, \Phi^{[2]}_{LL} (x_{1'2'}, x_{1'2'}, z') .
\end{align}

%%%%%%%%%%%%%%%%%%%%%%%%%%%%%%%%%%%%%%%%%%

\subsection{Type-2 contributions}
\label{sec:dsal_2}

The quark type-2 contribution is once again PT-odd and, as can be shown, vanishes. For the gluon type-2 operator, the calculation proceeds exactly as in the transverse case, with $\Phi^{\prime [2]}_{TT}$ and $\Phi^{[2]}_{T,-T}$ replaced by $\Phi^{[2]}_{TL}$ in the amplitude and $\Phi^{[2]}_{LT}$ replaced by $\Phi^{[2]}_{LL}$ in the complex conjugate amplitude. Inserting \eqs{A_eik3} and \eqref{A_subeik_2G} into \eq{dsal_lt}, using the wavefunction overlaps in \eqs{type2_overlaps}, and summing over $\lambda$, gives
\begin{align} \label{lt2G_l_1}
& \half \sum_{S_L = \pm 1} S_L \sum_{\lambda = \pm 1} \left[ e^{i \lambda \phi} \frac{d\sigma_{0 \lambda; \lambda_V = 0}^{\gamma^* p \to V p' \, \mathrm{G}[2]}}{d^2 \Delta} + \cc \right] = \frac{N_c \sqrt{2}}{4 \pi^2 \, s} \, \int \frac{d^2 x_1 \, d^2 x_2 \, d^2 x_0 \, d^2 x_{1'} \, d^2 x_{2'}}{(4 \pi)^2} \int\limits_0^1 \frac{dz}{z \, (1-z)} \int\limits_0^1 \frac{dz'}{z' \, (1-z')} \notag \\
& \times e^{- i \un \Delta \cdot \left[ z \un x_{1} + (1-z) \un x_2 - z' \un x_{1'} - (1-z')\, \un x_{2'} \right]} \, \Bigg\{ \sin (\phi - \phi_{12}) \, \Phi^{[2]}_{TL} (x_{12}, x_{12}, z) \, N_{12} (s) \Bigg[ \frac{1}{z'} \, \Phi^{[2]}_{LL} (x_{02'}, x_{1'2'}, z') \, \llangle \tord \tr \left[ V^{\textrm{G} [2]}_{\un 1', \un 0} V_{\un 2'}^\dagger \right] \rrangle^* (z' s) \notag \\
& \hspace{4.4cm} - \frac{1}{1-z'} \, \Phi^{[2]}_{LL} (x_{1'0}, x_{1'2'}, z') \, \llangle \tord \tr \left[ V_{\un 1'} V^{\textrm{G} [2] \, \dagger}_{\un 2', \un 0} \right] \rrangle^* ((1-z') s) \Bigg] \notag \\
& \hspace{0.4cm} + \Phi^{[2]}_{LL} (x_{1'2'}, x_{1'2'}, z') \, N_{1'2'} (s) \Bigg[ \frac{1}{z} \, \sin (\phi - \phi_{02}) \, \Phi^{[2]}_{TL} (x_{02}, x_{12}, z) \, \llangle \tord \tr \left[ V^{\textrm{G} [2]}_{\un 1, \un 0} V_{\un 2}^\dagger \right] \rrangle (z s) \notag \\
& \hspace{3.6cm} - \frac{1}{1-z} \, \sin (\phi - \phi_{10}) \, \Phi^{[2]}_{TL} (x_{10}, x_{12}, z) \, \llangle \tord \tr \left[ V_{\un 1} V^{\textrm{G} [2] \, \dagger}_{\un 2, \un 0} \right] \rrangle ((1-z) s) \Bigg] \Bigg\} + \cc .
\end{align}
Employing \eq{VG2sub}, discarding the PT-odd terms, adding the complex conjugate, and using the definition \eqref{Gi_def}, we find
\begin{align} \label{lt2G_l_2}
& \half \sum_{S_L = \pm 1} S_L \sum_{\lambda = \pm 1} \left[ e^{i \lambda \phi} \frac{d\sigma_{0 \lambda; \lambda_V = 0}^{\gamma^* p \to V p' \, \mathrm{G}[2]}}{d^2 \Delta} + \cc \right] = \frac{i \, N_c^2 \sqrt{2}}{\pi^2 \, s} \, \int \frac{d^2 x_1 \, d^2 x_2 \, d^2 x_{1'} \, d^2 x_{2'}}{(4 \pi)^2} \int\limits_0^1 \frac{dz}{z \, (1-z)} \int\limits_0^1 \frac{dz'}{z' \, (1-z')} \, \frac{1}{z} \notag \\
& \times e^{- i \un \Delta \cdot \left[ z \un x_{1} + (1-z) \un x_2 - z' \un x_{1'} - (1-z')\, \un x_{2'} \right]}  \, N_{1'2'} (s) \, G^i_{12} (z s) \notag \\
& \times \Bigg\{ \sin (\phi - \phi_{1'2'}) \, \Phi^{[2]}_{TL} (x_{1'2'}, x_{1'2'}, z')  \left[ \hat{x}^i_{12} \left( \Phi^{[2] (01)}_{LL} (x_{12}, x_{12}, z) - \Phi^{[2] (10)}_{LL} (x_{12}, x_{12}, z) \right) - i z \, \Delta^i \, \Phi^{[2]}_{LL} (x_{12}, x_{12}, z) \right] \notag \\
& \hspace{0.6cm} + \Phi^{[2]}_{LL} (x_{1'2'}, x_{1'2'}, z') \bigg[ \sin (\phi - \phi_{12}) \left( \hat{x}^i_{12} \left( \Phi^{[2] (01)}_{TL} (x_{12}, x_{12}, z) - \Phi^{[2] (10)}_{TL} (x_{12}, x_{12}, z) \right) - i z \, \Delta^i \, \Phi^{[2]}_{TL} (x_{12}, x_{12}, z) \right) \notag \\
& \hspace{3.6cm} - \cos (\phi - \phi_{12}) \, \frac{\epsilon^{ij} \, x_{12}^j}{x_{12}^2} \, \Phi^{[2]}_{TL} (x_{12}, x_{12}, z) \bigg] \Bigg\} .
\end{align}

%%%%%%%%%%%%%%%%%%%%%%%%%%%%%%%%%%%%%%%%%%%

\subsection{Type-3 contribution}
\label{sec:dsal_3}

Inserting \eqs{AeikF+-7} and \eqref{A_eik3} into \eq{dsal_lt}, using the TL and LL wavefunction overlaps from \eqs{type2_overlaps} and \eqref{type3_overlaps}, and summing over $\lambda$, we find
\begin{align} \label{lt3_l_raw}
& \half \sum_{S_L = \pm 1} S_L \sum_{\lambda = \pm 1} \left[ e^{i \lambda \phi} \frac{d\sigma_{0 \lambda; \lambda_V = 0}^{\gamma^* p \to V p' \, [3]}}{d^2 \Delta} + \cc \right] = - \frac{i \, N_c \sqrt{2}}{4 \pi^2 \, s} \, \int \frac{d^2 x_1 \, d^2 x_2 \, d^2 x_{1'} \, d^2 x_{2'}}{(4 \pi)^2} \int\limits_0^1 \frac{dz}{z \, (1-z)} \int\limits_0^1 \frac{dz'}{z' \, (1-z')} \notag \\
& \times e^{- i \un \Delta \cdot \left[ z \un x_{1} + (1-z) \un x_2 - z' \un x_{1'} - (1-z')\, \un x_{2'} \right]}  \, \sin (\phi - \phi_{12}) \, \Bigg\{ \Phi^{[2]}_{TL} (x_{12}, x_{12}, z) \left[ \Phi^{[3]}_{LL} (x_{1'2'}, z') + \frac{z' - \thalf}{z' (1-z')} \, \Phi^{[2]}_{LL} (x_{1'2'}, x_{1'2'}, z') \right] \notag \\
& \hspace{2.2cm} \times N_{12}(s) \, \left[ \llangle \atord \tr \left[ V_{\un 2'} V^{\textrm{G} [3] \, \dagger}_{\un 1'} \right] \rrangle (z's) - \llangle \atord \tr \left[ V^{\textrm{G} [3]}_{\un 2'} V^\dagger_{\un 1'} \right] \rrangle ((1-z') s) \right] \notag \\
& \hspace{0.6cm} - \Phi^{[2]}_{LL} (x_{1'2'}, x_{1'2'}, z') \left[ \Phi^{[3]}_{TL} (x_{12}, z) + \frac{z - \thalf}{z (1-z)} \, \Phi^{[2]}_{TL} (x_{12}, x_{12}, z) \right] \notag \\
& \hspace{2.2cm} \times N_{1'2'}(s) \, \left[ \llangle \tord \tr \left[ V^{\textrm{G} [3]}_{\un 1} V^\dagger_{\un 2} \right] \rrangle (zs) - \llangle \tord \tr \left[ V_{\un 1} V^{\textrm{G} [3] \, \dagger}_{\un 2} \right] \rrangle ((1-z) s) \right] \Bigg\} + \cc .
\end{align}
Interchanging $\un x_{1'} \leftrightarrow \un x_{2'}$, $z' \leftrightarrow 1-z'$ in the trace involving $V^{\mathrm{G}[3]}_{\un 2'}$ of \eq{lt3_l_raw} and $\un x_{1} \leftrightarrow \un x_{2}$, $z \leftrightarrow 1-z$ in the trace involving $V^{\mathrm{G}[3]}_{\un 2}$
turns each difference of traces into a sum, exactly as in the transverse case. Adding the complex conjugate and using \eq{G3_def}, we arrive at
\begin{align} \label{lt3_l_final}
& \half \sum_{S_L = \pm 1} S_L \sum_{\lambda = \pm 1} \left[ e^{i \lambda \phi} \frac{d\sigma_{0 \lambda; \lambda_V = 0}^{\gamma^* p \to V p' \, [3]}}{d^2 \Delta} + \cc \right] = \frac{i \, N_c^2 \sqrt{2}}{\pi^2 \, s} \, \int \frac{d^2 x_1 \, d^2 x_2 \, d^2 x_{1'} \, d^2 x_{2'}}{(4 \pi)^2} \int\limits_0^1 \frac{dz}{z \, (1-z)} \int\limits_0^1 \frac{dz'}{z' \, (1-z')} \notag \\
& \times e^{- i \un \Delta \cdot \left[ z \un x_{1} + (1-z) \un x_2 - z' \un x_{1'} - (1-z')\, \un x_{2'} \right]}  \, N_{1'2'}(s) \, G^{[3]}_{12} (zs) 
\notag \\ & \times 
\Bigg\{ \sin (\phi - \phi_{1'2'}) \, \Phi^{[2]}_{TL} (x_{1'2'}, x_{1'2'}, z') \left[ \Phi^{[3]}_{LL} (x_{12}, z) + \frac{z - \thalf}{z (1-z)} \, \Phi^{[2]}_{LL} (x_{12}, x_{12}, z) \right] \notag \\
& \hspace{4.6cm} + \sin (\phi - \phi_{12}) \, \Phi^{[2]}_{LL} (x_{1'2'}, x_{1'2'}, z') \left[ \Phi^{[3]}_{TL} (x_{12}, z) + \frac{z - \thalf}{z (1-z)} \, \Phi^{[2]}_{TL} (x_{12}, x_{12}, z) \right] \Bigg\} .
\end{align}
The complete LT structure for a longitudinally polarized vector meson is the sum of \eqs{lt1_l}, \eqref{lt2G_l_2}, and \eqref{lt3_l_final}.

\subsection{Expansion in transverse momentum transfer $\Delta_\perp$}
\label{sec:dsal_expansion}

We expand \eqs{lt1_l}, \eqref{lt2G_l_2}, and \eqref{lt3_l_final} to linear order in $\Delta_\perp$, exactly as in Sec.~\ref{sec:dsat_expansion}. After integrating over the impact parameters via \eqs{int_defs}, \eqref{Nint}, and \eqref{b_ints} and over the dipole orientations $\phi_{12}$ and $\phi_{1'2'}$, the result again consists of two harminics: $\un \Delta \cdot \hat k$ and $\un \Delta \times \hat k$. The parity argument given in Sec.~\ref{sec:dsat_expansion} applies here as well, so we expect the $\un \Delta \times \hat{k}$ harmonic to vanish identically. Isolating the $\cos (\phi - \phi_\Delta)$ harmonic gives our main result for the longitudinally polarized vector mesons,
\begin{tcolorbox}
\begin{align} \label{lt_exp_l_cos}
& \half \sum_{S_L = \pm 1} S_L \sum_{\lambda = \pm 1} \left[ e^{i \lambda \phi} \frac{d\sigma_{0 \lambda; \lambda_V = 0}^{\gamma^* p \to V p'}}{d^2 \Delta} + \cc \right]_{\cos (\phi - \phi_\Delta)} \notag \\
& = \un \Delta \cdot \hat{k} \; \frac{2 \, N_c^2 \sqrt{2}}{\pi^2 \, s} \int \frac{d x^2_{12} \, d x^2_{1'2'}}{64} \int\limits_0^1 \frac{dz}{z \, (1-z)} \int\limits_0^1 \frac{dz'}{z' \, (1-z')} \, \frac{1}{z} \, N (x^2_{1'2'} , s) \, \Phi^{[2]}_{LL} (x_{1'2'}, x_{1'2'}, z') \notag \\
& \times \Bigg\{ x_{12} \, \Phi^{[1]}_{TL} (x_{12}, z) \Big( I_3 (x_{12}^2, zs) - (1-z) \, Q (x_{12}^2, zs) \Big) \notag \\
& \hspace{0.6cm} - x_{12}^2 \left( \Phi^{[2] (01)}_{TL} (x_{12}, x_{12}, z) - \Phi^{[2] (10)}_{TL} (x_{12}, x_{12}, z) \right) \Big( I_4 (x_{12}^2, zs) + I_5 (x_{12}^2, zs) \Big) \notag \\
& \hspace{0.6cm} + x_{12} \, \Phi^{[2]}_{TL} (x_{12}, x_{12}, z) \Big( I_4 (x_{12}^2, zs) + (1-2z) \, G_2 (x_{12}^2, zs) \Big) \notag \\
& \hspace{0.6cm} - z \, x_{12} \left( \Phi^{[3]}_{TL} (x_{12}, z) + \frac{z - \thalf}{z (1-z)} \, \Phi^{[2]}_{TL} (x_{12}, x_{12}, z) \right) I_E (x_{12}^2, zs) \Bigg\} + {\cal O} (\Delta_\perp^2) ,
\end{align}
\end{tcolorbox}
\noindent with the explicit expressions for the wavefunction overlaps given in Appendix~\ref{app:overlaps}. 

As in the case of the transverse polarization for the vector meson, every ingredient of \eq{lt_exp_l_cos} that is not a moment amplitude is known. The amplitudes $Q$ and $G_2$ are constrained by the small-$x$ helicity phenomenology \cite{Adamiak:2021ppq, Adamiak:2023yhz, JAMCollaborationSmall-xAnalysisGroup:2025tfa}, while $N$ is fixed by the unpolarized data collected by many accelerators in various types of collisions. Comparing \eq{lt_exp_l_cos} with \eq{lt_exp_cos}, one sees that the same moment amplitudes $I_3$, $I_4$, $I_5$, and $I_E$ appear in both expressions, but weighted by different light-cone wave function overlaps. These overlaps have different dependence on $Q^2$, $z$, and the dipole sizes. Measuring the two meson polarizations separately therefore delivers two independent linear combinations of the same underlying moment amplitudes. We now can conclude that the measurement of the $\cos (\phi - \phi_\Delta)$ harmonic in the DSA for the elastic production of a transversely or longitudinally polarized vector meson at the EIC may help constrain the gluon OAM distribution in \eq{final_gluon_OAM}.

%%%%%%%%%%%%%%%%%%%%%%%%%%%%%%%%%%%%%%%%%%%%%%%%%%%%%%%%%%%%%%%%%%%%%%%%%%%%%%%

\section{Conclusion and outlook}
\label{sec:conclusions}

In this work we have computed the DSA for elastic vector meson production in the scattering of a longitudinally polarized electron on a longitudinally polarized proton at small $x$. Working in the LCOT formalism at sub-eikonal order, we included all three types of polarized Wilson line operators, namely the type-1 operators $V^{\mathrm{G}[1]}$ and $V^{\mathrm{q}[1]}$, the type-2 operators $V^{\mathrm{G}[2]}$ and $V^{\mathrm{q}[2]}$, and the type-3 operator $V^{\mathrm{G}[3]}$, which arises from the $x^+$-dependence of the Wilson lines. The type-2 quark operator drops out of the DSA because its contribution is PT-odd, while the remaining operators give a non-zero result.  

Our main results are \eqs{lt_exp_cos} and \eqref{lt_exp_l_cos}. The OAM sensitivity resides in the interference between longitudinally and transversely polarized virtual photons, and it appears at linear order in the momentum transfer $\Delta_\perp$, in the $\cos(\phi - \phi_\Delta)$ harmonic of the DSA. For transversely and longitudinally polarized mesons alike, this harmonic is a linear combination of the moment amplitudes $I_3$, $I_4$, $I_5$, and $I_E$, which govern the small-$x$ OAM distributions in \eqs{OAMs_corr}. The accompanying terms involve only $Q$ and $G_2$, already constrained by small-$x$ helicity phenomenology, and the unpolarized dipole amplitude $N$. Because the two meson polarizations weight the moment amplitudes with different wave function overlaps, separating them experimentally through the angular distribution of the vector meson decay products would provide two independent constraints on the moment-amplitudes. Furthermore, since different mesons have different light-cone wavefunctions, our results give two independent OAM-sensitive observables for every species of vector meson that is measured. 

Elastic vector meson production thus joins elastic dijet production \cite{Kovchegov:2024wjs} as an exclusive channel with direct access to OAM at small $x$. The two processes are complementary. Dijet production offers an additional handle through the relative transverse momentum of the jets, while vector meson production benefits from a clean final state and an extensive body of unpolarized measurements at HERA. A combined analysis of both channels should considerably improve the constraining power on the moment amplitudes.

We also emphasize our results here can be also be applied to deeply-virtual Compton scattering (DVCS)\footnote{We would like to thank Simonetta Liuti for a discussion on this subject.} by replacing the vector meson light-cone wavefunction throughout this paper with the (real) photon one. The vector meson wave function is a non-perturbative object which has to be modeled to perform phenomenological analyses with our results: this weakness of our presently considered vector meson channel is remedied in DVCS, since the wavefunction for the photon splitting into a quark--anti-quark pair is perturbative (in QED and, for smaller dipoles, in QCD) and is well-known \cite{Hanninen:2017ddy}. In the case of DVCS, however, one would also need to consider the photon-mediated Bethe-Heitler process in addition to the QCD-mediated scattering considered in this work. We therefore leave a more complete treatment of DVCS for future work. 

Several other directions also remain for future work. The most immediate is a quantitative feasibility study. One would combine the small-$x$ helicity extractions of the JAM collaboration \cite{Adamiak:2021ppq, Adamiak:2023yhz, JAMCollaborationSmall-xAnalysisGroup:2025tfa} with the small-$x$ evolution of the moment amplitudes \cite{Kovchegov:2023yzd, Becker:2026xcp} and generate EIC pseudo-data for the $\cos(\phi - \phi_\Delta)$ harmonic of $A_{LL}$, for light ($\rho$, $\phi$) as well as heavy ($J/\psi$) vector mesons. Such a study would also have to quantify the model dependence associated with the vector meson light-cone wave functions, for which we have used the standard photon-like ansatz \cite{Kowalski:2003hm, Nemchik:1994fp, Nemchik:1996cw, Kowalski:2006hc}. Furthermore, it would be interesting to see how corrections beyond the DLA approximation employed here affect the feasibility of elastic vector meson production as a probe for the OAM distributions.

%%%%%%%%%%%%%%%%%%%%%%%%%%%%%%%%%%%%%%%%%%%%%%%%%%%%%%%%%%%%%%%%%%%%%%%%%%%%%%%

\section*{Acknowledgments}

The authors are grateful to Shohini Bhattacharya, Daniel Brandenburg, Adrian Dumitru, Yoshitaka Hatta, Simonetta Liuti, and Daniel Pitonyak for informative discussions. 

This material is based upon work supported by the U.S. Department of Energy, Office of Science, Office of Nuclear Physics under Award Number DE-SC0004286, by the U.S. National Science Foundation under Grant No.~PHY-2609761 (BM), and within the framework of the Saturated Glue (SURGE) Topical Theory Collaboration.
\\

%%%%%%%%%%%%%%%%%%%%%%%%%%%%%%%%%%%%%%%%%%%%%%%%%%%%%%%%%%%%%%%%%%%%%%%%%%%%%%%

\appendix

\section{Explicit expressions for the wave function overlaps}
\label{app:overlaps}

For convenience we collect here the explicit forms of every wave function overlap entering the results of Sections~\ref{sec:dsa_t} and \ref{sec:dsa_l}. They follow from the light-cone wavefunctions of the virtual photon and of the vector meson \cite{Kowalski:2003hm, Nemchik:1994fp, Nemchik:1996cw, Kowalski:2006hc} through the definitions in \eqs{type1_overlaps}, \eqref{type2_overlaps}, and \eqref{type3_overlaps}. Here $\phi_T$ and $\phi_L$ are the scalar parts of the transverse and longitudinal vector meson wavefunctions, $M_V$ is the meson mass, $\delta = 0, 1$ switches off and on the derivative term of the boosted Gaussian model, and
\begin{align} \label{app_af}
a_f = \sqrt{Q^2 \, z (1-z) + m_f^2}
\end{align}
with $m_f$ the mass of the quark of flavor $f$. A sum over the quark flavors making up the meson, weighted by the appropriate flavor coefficients, is implicit throughout.

The type-1 overlaps defined in \eqs{type1_overlaps} read
\begin{subequations} \label{app_type1}
\begin{align}
\Phi^{[1]}_{TT} (x_\perp, z) &= \frac{e Z_f}{\pi} \, (1-2z) \, a_f \, K_1 (x_\perp \, a_f) \, \frac{\pd \phi_T (x_\perp, z)}{\pd x_\perp} , \\
\Phi^{\prime \, [1]}_{TT} (x_\perp, z) &= \frac{e Z_f}{\pi} \, m_f^2 \, K_0 (x_\perp \, a_f) \, \phi_T (x_\perp, z) , \\
\Phi^{[1]}_{TL} (x_\perp, z) &= \frac{e Z_f}{\pi} \, z (1-z) \, a_f \, K_1 (x_\perp \, a_f) \left[ M_V + \delta \, \frac{m_f^2 - \nabla_\perp^2}{M_V \, z (1-z)} \right] \phi_L (x_\perp, z) , \\
\Phi^{[1]}_{LT} (x_\perp, z) &= \frac{e Z_f}{\pi} \, z (1-z) \, 2 Q \, K_0 (x_\perp \, a_f) \, \frac{\pd \phi_T (x_\perp, z)}{\pd x_\perp} .
\end{align}
\end{subequations}

The type-2 overlaps of \eqs{type2_overlaps} carry two independent transverse arguments, the first belonging to the photon wave function and the second to the meson wave function. The Bessel functions are therefore always evaluated at the first argument, and the meson wave functions at the second. In the notation of \eq{Phi_decomp} they are
\begin{subequations} \label{app_type2}
\begin{align}
\Phi^{\prime [2A]}_{TT} (x_\perp, y_\perp, z) &= - \frac{e Z_f}{2 \pi} \, \frac{1 + (1-2z)^2}{x_\perp \, y_\perp} \, a_f \, K_1 (x_\perp \, a_f) \, \frac{\pd \phi_T (y_\perp, z)}{\pd y_\perp} , \\
\Phi^{\prime [2B]}_{TT} (x_\perp, y_\perp, z) &= \frac{e Z_f}{\pi} \, m_f^2 \, K_0 (x_\perp \, a_f) \, \phi_T (y_\perp, z) , \\
\Phi^{[2]}_{T,-T} (x_\perp, y_\perp, z) &= - \frac{2 \, e Z_f}{\pi} \, z (1-z) \, a_f \, K_1 (x_\perp \, a_f) \, \frac{\pd \phi_T (y_\perp, z)}{\pd y_\perp} , \\
\Phi^{[2]}_{TL} (x_\perp, y_\perp, z) &= - \frac{e Z_f}{\pi} \, z (1-z) (1-2z) \, a_f \, K_1 (x_\perp \, a_f) \left[ M_V + \delta \, \frac{m_f^2 - \nabla_\perp^2}{M_V \, z (1-z)} \right] \phi_L (y_\perp, z) , \\
\Phi^{[2]}_{LT} (x_\perp, y_\perp, z) &= - \frac{e Z_f}{\pi} \, z (1-z) (1-2z) \, 2 Q \, K_0 (x_\perp \, a_f) \, \frac{\pd \phi_T (y_\perp, z)}{\pd y_\perp} , \\
\Phi^{[2]}_{LL} (x_\perp, y_\perp, z) &= \frac{e Z_f}{\pi} \, z^2 (1-z)^2 \, 2 Q \, K_0 (x_\perp \, a_f) \left[ M_V + \delta \, \frac{m_f^2 - \nabla_\perp^2}{M_V \, z (1-z)} \right] \phi_L (y_\perp, z) .
\end{align}
\end{subequations}
The function $\overline{\Phi}^{[2]}_{TT}$ of \eq{Phi_decomp} is not independent,
\begin{align} \label{app_Phibar}
\overline{\Phi}^{[2]}_{TT} (x_\perp, y_\perp, z) = \Phi^{\prime [2A]}_{TT} (x_\perp, y_\perp, z) ,
\end{align}
since the dot product and the cross product in \eq{TTT-T} multiply the same scalar function. At coincident arguments the TL and LT overlaps reduce to their type-1 counterparts,
\begin{align} \label{app_coincident}
\Phi^{[2]}_{TL} (x_\perp, x_\perp, z) = - (1-2z) \, \Phi^{[1]}_{TL} (x_\perp, z) , \qquad \Phi^{[2]}_{LT} (x_\perp, x_\perp, z) = - (1-2z) \, \Phi^{[1]}_{LT} (x_\perp, z) .
\end{align}

The type-2 gluon operator also generates the derivative combinations defined in \eq{deriv_not}. Only the antisymmetric combination $\Phi^{(01)} - \Phi^{(10)}$ enters the results, which gives
\begin{subequations} \label{app_derivs}
\begin{align}
& x_\perp^2 \left[ \Phi^{\prime [2A] (01)}_{TT} - \Phi^{\prime [2A] (10)}_{TT} \right] (x_\perp, x_\perp, z) = - \frac{e Z_f}{2 \pi} \left[ 1 + (1-2z)^2 \right] \Bigg[ a_f \, K_1 (x_\perp \, a_f) \, \frac{\pd^2 \phi_T (x_\perp, z)}{\pd x_\perp^2} \notag \\
& \hspace{7.4cm} + \left( a_f^2 \, K_0 (x_\perp \, a_f) + \frac{a_f \, K_1 (x_\perp \, a_f)}{x_\perp} \right) \frac{\pd \phi_T (x_\perp, z)}{\pd x_\perp} \Bigg] , \\
& \left[ \Phi^{\prime [2B] (01)}_{TT} - \Phi^{\prime [2B] (10)}_{TT} \right] (x_\perp, x_\perp, z) = \frac{e Z_f}{\pi} \, m_f^2 \Bigg[ K_0 (x_\perp \, a_f) \, \frac{\pd \phi_T (x_\perp, z)}{\pd x_\perp} \notag \\
& \hspace{7.4cm} + a_f \, K_1 (x_\perp \, a_f) \, \phi_T (x_\perp, z) \Bigg] , \\
& \left[ \Phi^{[2] (01)}_{T,-T} - \Phi^{[2] (10)}_{T,-T} \right] (x_\perp, x_\perp, z) = - \frac{2 \, e Z_f}{\pi} \, z (1-z) \Bigg[ a_f \, K_1 (x_\perp \, a_f) \, \frac{\pd^2 \phi_T (x_\perp, z)}{\pd x_\perp^2} \notag \\
& \hspace{7.0cm} + \left( a_f^2 \, K_0 (x_\perp \, a_f) + \frac{a_f \, K_1 (x_\perp \, a_f)}{x_\perp} \right) \frac{\pd \phi_T (x_\perp, z)}{\pd x_\perp} \Bigg] , \\
& \left[ \Phi^{[2] (01)}_{TL} - \Phi^{[2] (10)}_{TL} \right] (x_\perp, x_\perp, z) = - \frac{e Z_f}{\pi} \, z (1-z) (1-2z) \notag \\
& \hspace{4.6cm} \times \left[ a_f \, K_1 (x_\perp \, a_f) \, \frac{\pd}{\pd x_\perp} + a_f^2 \, K_0 (x_\perp \, a_f) + \frac{a_f \, K_1 (x_\perp \, a_f)}{x_\perp} \right] \notag \\
& \hspace{6.4cm} \times \left[ M_V + \delta \, \frac{m_f^2 - \nabla_\perp^2}{M_V \, z (1-z)} \right] \phi_L (x_\perp, z) , \\
& \left[ \Phi^{[2] (01)}_{LT} - \Phi^{[2] (10)}_{LT} \right] (x_\perp, x_\perp, z) = - \frac{e Z_f}{\pi} \, z (1-z) (1-2z) \, 2 Q \Bigg[ K_0 (x_\perp \, a_f) \, \frac{\pd^2 \phi_T (x_\perp, z)}{\pd x_\perp^2} \notag \\
& \hspace{7.4cm} + a_f \, K_1 (x_\perp \, a_f) \, \frac{\pd \phi_T (x_\perp, z)}{\pd x_\perp} \Bigg] , \\
& \left[ \Phi^{[2] (01)}_{LL} - \Phi^{[2] (10)}_{LL} \right] (x_\perp, x_\perp, z) = \frac{e Z_f}{\pi} \, z^2 (1-z)^2 \, 2 Q \left[ K_0 (x_\perp \, a_f) \, \frac{\pd}{\pd x_\perp} + a_f \, K_1 (x_\perp \, a_f) \right] \notag \\
& \hspace{6.4cm} \times \left[ M_V + \delta \, \frac{m_f^2 - \nabla_\perp^2}{M_V \, z (1-z)} \right] \phi_L (x_\perp, z) .
\end{align}
\end{subequations}
In the two expressions involving $\phi_L$ the derivative operator in the first square brackets acts on everything to its right.

Finally, the overlaps generated by the $z$-derivative of the photon wave function, defined in \eqs{type3_overlaps}, are
\begin{subequations} \label{app_type3}
\begin{align}
\Phi^{\prime \, [3]}_{TT} (x_\perp, z) &= \frac{e Z_f}{\pi} \, (1-2z) \Bigg\{ \left[ \frac{1 + (1-2z)^2}{4} \, x_\perp \, Q^2 \, K_0 (x_\perp \, a_f) - \frac{(1-2z)^2}{2 z (1-z)} \, a_f \, K_1 (x_\perp \, a_f) \right] \frac{\pd \phi_T (x_\perp, z)}{\pd x_\perp} \notag \\
& \hspace{2.2cm} + m_f^2 \left[ \frac{K_0 (x_\perp \, a_f)}{2 z (1-z)} - \frac{x_\perp \, Q^2}{2 \, a_f} \, K_1 (x_\perp \, a_f) \right] \phi_T (x_\perp, z) \Bigg\} , \\
\Phi^{[3]}_{T,-T} (x_\perp, z) &= \frac{e Z_f}{\pi} \, (1-2z) \left[ z (1-z) \, x_\perp \, Q^2 \, K_0 (x_\perp \, a_f) - 2 \, a_f \, K_1 (x_\perp \, a_f) \right] \frac{\pd \phi_T (x_\perp, z)}{\pd x_\perp} , \\
\Phi^{[3]}_{TL} (x_\perp, z) &= - \frac{e Z_f}{2 \pi} \left[ \big( 1 - 8 z (1-z) \big) \, a_f \, K_1 (x_\perp \, a_f) - z (1-z) (1-2z)^2 \, x_\perp \, Q^2 \, K_0 (x_\perp \, a_f) \right] \notag \\
& \hspace{1.0cm} \times \left[ M_V + \delta \, \frac{m_f^2 - \nabla_\perp^2}{M_V \, z (1-z)} \right] \phi_L (x_\perp, z) , \\
\Phi^{[3]}_{LT} (x_\perp, z) &= \frac{e Z_f}{\pi} \, Q \, (1-2z)^2 \left[ 3 \, K_0 (x_\perp \, a_f) - \frac{z (1-z) \, x_\perp \, Q^2}{a_f} \, K_1 (x_\perp \, a_f) \right] \frac{\pd \phi_T (x_\perp, z)}{\pd x_\perp} , \\
\Phi^{[3]}_{LL} (x_\perp, z) &= \frac{e Z_f}{\pi} \, Q \, z (1-z) (1-2z) \left[ 3 \, K_0 (x_\perp \, a_f) - \frac{z (1-z) \, x_\perp \, Q^2}{a_f} \, K_1 (x_\perp \, a_f) \right] \notag \\
& \hspace{1.0cm} \times \left[ M_V + \delta \, \frac{m_f^2 - \nabla_\perp^2}{M_V \, z (1-z)} \right] \phi_L (x_\perp, z) .
\end{align}
\end{subequations}

Once the model functions $\phi_{T,L}(x_\perp, z)$ are specified for the vector meson wavefunction, the wavefunction overlaps can be computed via the formulas above for use in \eqs{lt_exp_cos} and \eqref{lt_exp_l_cos}.

%%%%%%%%%%%%%%%%%%%%%%%%%%%%%%%%%%%%%%%%%%%%%%%%%%%%%%%%%%%%%%%%%%%%%%%%%%%%%%%

\providecommand{\href}[2]{#2}\begingroup\raggedright\endgroup

\end{document}